\documentclass[letterpaper]{article}
\pdfoutput=1
\usepackage{jheppub}

\usepackage[utf8]{inputenc}
\usepackage[english]{babel}
\usepackage{amsmath}
\usepackage{amsfonts}
\usepackage{amssymb}
\usepackage{listings}
\usepackage{lipsum}
\usepackage{multirow}
\usepackage{datetime}
\usepackage{graphicx}
\usepackage{mathtools}
\usepackage{mathrsfs}
\usepackage{dcolumn}
\usepackage{multirow}
\usepackage{color}   
\usepackage{hyperref}
\hypersetup{
    colorlinks=true, 
    pdfborder = {0 0 0.5 [3 3]},
    anchorcolor=black,
    citecolor=blue,
    linktoc=all,    
    linktocpage=true,
    linkcolor=red,
	urlcolor=blue
}

\newcommand{\vect}[1]{\boldsymbol{\mathbf{#1}}}

\title{Dark photon vortex formation and dynamics}

\author{William E.\ East,}
\author{Junwu Huang}

\affiliation{Perimeter Institute for Theoretical Physics, Waterloo, Ontario N2L 2Y5, Canada}
\emailAdd{weast@perimeterinstitute.ca}
\emailAdd{jhuang@perimeterinstitute.ca}
\date{\today}

\abstract{ We study the formation and evolution of vortices in $U(1)$ dark
photon dark matter and dark photon clouds that arise through black hole
superradiance. We show how the production of both longitudinal mode and transverse mode dark photon dark matter 
can lead to the formation of vortices. After vortex formation, the energy stored in
the dark photon dark matter will be transformed into a large number of vortex
strings, eradicating the coherent dark photon dark matter field. In the case where a dark photon magnetic field is produced, bundles of vortex strings are formed in a superheated phase
transition, and evolve towards a configuration consisting of many string loops that are uncorrelated on large scales,
analogous to a melting phase transition in condensed matter.
In the process, they dissipate via dark photon and gravitational wave emission, 
offering a target for experimental searches. 
Vortex strings were also recently shown to form in dark photon superradiance clouds around black holes,
and we discuss the dynamics and observational consequences of this phenomenon with phenomenologically
motivated parameters. In that case, the string loops ejected from the
superradiance cloud, apart from producing gravitational waves, are also quantised magnetic flux lines and can be looked for
with magnetometers. We discuss the connection 
between the dynamics in these scenarios and similar vortex dynamics found
in type II superconductors.
}

\begin{document}

\maketitle

\section{Introduction and summary}\label{sec:intro}

The dark photon~\cite{Holdom:1985ag} is a $U(1)$ vector boson which has
been studied extensively as a candidate for new physics beyond
the Standard Model. Motivated by string
theory~\cite{Arvanitaki:2009fg,Goodsell:2009xc}, the light vector field
can play a significant role in dark matter direct
detection~\cite{Pospelov:2008jk,Essig:2011nj},
stellar~\cite{An:2013yfc,Hardy:2016kme}, galactic~\cite{Lasenby:2020rlf},
and cosmological~\cite{Jaeckel:2008fi,Caputo:2020bdy} dynamics. The
bosonic nature of the dark photon allows it to be produced with very large
amplitude in clouds around black holes that arise through superradiance~\cite{Baryakhtar:2017ngi}, 
and by inflationary perturbations as well
as parametric resonance into dark photon dark
matter~\cite{Graham:2015rva,Nelson:2011sf,Agrawal:2018vin}. Dark
photon dark matter, in particular, is a prime target of a number of
experiments~\cite{Chaudhuri:2014dla,PhysRevD.98.035006,PhysRevD.101.052008,PhysRevD.102.042001,PhysRevLett.125.171802,Chiles:2021gxk,Cervantes:2022yzp}.

At low energies and small amplitudes, the dark photon is described by the Proca action
\begin{equation}
    \mathcal{S} = \int {\rm d}^4 x \left( -\frac{1}{4}F'^{\mu\nu}F'_{\mu\nu} -\frac{1}{2} m_{A'}^2 A'_{\mu}A'^{\mu}\right),
\end{equation}
which describes a massive vector field $A'^{\mu}$ with mass $m_{A'}$ and field
tensor $F'_{\mu\nu}=\partial_\mu A'_{\nu}-\partial_\nu A'_{\mu}$. The Abelian Higgs model
provides a UV completion of the Proca field above the scale $m_{A'}$, and has the
following action
\begin{equation}
    \mathcal{S} = \int {\rm d}^4 x  \left[\frac{1}{2}\left|D'_{\mu} \Phi\right|^2 -\frac{1}{4}F'^{\mu\nu}F'_{\mu\nu}-\frac{\lambda}{4}\left(|\Phi|^2-v^2\right)^2\right],
\end{equation}\label{eq:abelianhiggs}
where $\Phi$ is complex scalar, $D'_{\mu} \equiv \partial_{\mu} - i g_D A'_{\mu}$ is the covariant derivative,
and $g_D$ and $\lambda$ are coupling constants. We can write the complex scalar in terms 
of a phase and a radial displacement from its vacuum expectation value (VEV) $v$ as $\Phi = (v+\rho) \exp (i \Pi/v)$.
The mass of the radial mode is $m_\rho=\sqrt{2}\lambda^{1/2} v$, while the dark photon
mass is $m_{A'} = g_D v$.

Dark photon dark matter searches are based on the limit where the quartic
coupling $\lambda$ is taken to be infinite compared to the gauge coupling
$g_D$, and the radial mode of the scalar $\Phi$ is taken to be infinitely
heavy\footnote{This infinite $\lambda$ limit simply indicates that the radial
mode is likely a composite state of a complicated UV theory beyond the scale of
$\lambda^{1/2} v$~\cite{osti_7264935}, like in a high-$T_c$ superconductor.}.
Below the scale $\lambda^{1/2} v$, the radial mode can be integrated out, and
the dynamics of the system is assumed to be described by the Stueckelberg
action~\cite{2009esuf}
\begin{equation}
    \mathcal{S} = \int {\rm d}^4 x \left[ -\frac{1}{4}F'^{\mu\nu}F'_{\mu\nu} -\frac{1}{2} \left(m_{A'}A'_{\mu}-\partial_{\mu} \Pi\right)\left(m_{A'}A'^{\mu}-\partial^{\mu}\Pi \right)\right],
\end{equation}
which reduces to the Proca action in the unitary gauge ($\Pi=0$). This action is sometimes assumed to be valid as long as the symmetry is not restored at strong field (see~\cite{Lam:1971my,tinkham2004introduction} and Appendix \ref{sec:eomKG} for more details):
\begin{equation}
    g_D \rho_{A'}^{1/2} \leq g_D B_{\rm c2}' \equiv  \lambda v^2
\label{eq:condition1}
\end{equation}
where $\rho_{A'}=(E'^2+B'^2)/2$ is the energy density due to the dark electric and magnetic fields.

This picture, however, is incomplete. Rather, there are vortex like solutions
which are the lowest energy solution before the field strength reaches $B_{\rm
c2}'$ in eq.~\ref{eq:condition1}. In the limit of infinite $\lambda$, these
vortex solutions are described by the Nambu-Goto action,
integrated along the world sheet of the string vortex~\cite{NambuGoto}
\begin{equation}
     \mathcal{S} = - \mu \int {\rm d} t  {\rm d} l \sqrt{\gamma},
\end{equation}
where the string tension $\mu$ is $\mathcal{O}(v^2)$, with only a logarithmic
dependence on the ratio between the string core size and the size of the region
which contains the magnetic field [and hence up to $\log(\lambda/g_D^2)$].
These vortex solutions are very similar to the vortices that form in type II
superconductors, the well known Abrikosov
vortices~\cite{Abrikosov:1956sx} (for more details about this analogy and the definition of the variables, see table~\ref{tab:variabledefinitions}). In all cases we consider, we find that the vortices
form in background energy densities that are significantly smaller than
$B_{\rm c2}'^2 =\lambda^2v^4/g_D^2$.
In particular, they can form in a finite energy density in the limit
$g_D\rightarrow 0$. The vortex solution is the energetically favorable solution
in the background field as long as 
\begin{equation}
\rho_{A'}^{1/2} \gtrsim B_{\rm c1}' \approx g_D v^2
\end{equation}
when $\lambda\gg g_D^2$~\cite{Abrikosov:1956sx,Nielsen:1973cs}. 
Therefore, once formed, they will quickly dissipate energy stored in the dark photon dark
matter or superradiance cloud and completely change the dynamics of the system
as well as the phenomenological consequences. We will discuss in detail how
these vortices form, how they evolve in the background dark electric and magnetic field,
and the phenomenological consequences in the context of dark photon dark matter
and a dark photon superradiance cloud.

In this work, we argue that 
vortex production occurs in all proposed dark photon dark matter production
mechanisms~\cite{Graham:2015rva,Agrawal:2018vin,Co:2018lka,Dror:2018pdh,Bastero-Gil:2018uel}.
During inflation, dark photon dark matter can be produced as longitudinal modes. If
the Hubble parameter during inflation $H_I \gtrsim v$, regardless of if the $U(1)$ symmetry is restored or not, that
is, independent of if $\lambda \rightarrow \infty$ ($m_{\rho} \gg H_I$),
vortices will form. The evolution of these strings is qualitatively the same
as ordinary cosmic gauged strings (see Appendix \ref{sec:NonMCoupling} for
some discussions about the differences), which after inflation approach a
scaling solution. 

Vortex production can also happen in the late Universe when
energy is transferred from axions to the dark photon magnetic
field~\cite{Agrawal:2018vin,Co:2018lka,Dror:2018pdh,Bastero-Gil:2018uel}. In
this case, the phase transition is a superheated phase transition when the
magnetic field in the system reaches the superheating field 
\begin{equation}
    B_{\rm sh}' \equiv \lambda^{1/2} v^2.
\label{eqn:sh}
\end{equation}
In the context of superconductors, at the superheating field, the 
transition between a phase with uniform magnetic field, to a phase with a
large number of magnetic vortices, becomes a first order phase transition, though
the magnetic field strength is still much smaller than $B_{\rm c2}'$. 

Though in the above we have borrowed terminology from the superconductor literature,
the superheating field threshold, eq.~\ref{eqn:sh}, is parametrically the same as one would derive
from equating the vector field energy density to the vacuum energy density of the Higgs
$\rho_{A'}\sim \lambda v^4$. Alternatively, one could consider the equation of motion
for the radial mode, which is given by $\Box \rho = V_{\rm eff}'(\rho)$ where
\begin{equation}
    V_{\rm eff} = \frac{\lambda}{4}\left[(\rho+v)^2-v^2\right]^2+\frac{1}{2}g_D^2 X(\rho+v)^2
    \mathrm{\ with \ } X=\left(A'_\mu-\frac{1}{m_{A'}}\partial_\mu\Pi\right)^2 \ .
\label{eqn:veff}
\end{equation}
When $X\geq \lambda v^2/g_D^2$, the minimum
of $V_{\rm eff}$ moves to $\Phi=0$ ($\rho=-v$)~\cite{Fukuda:2019ewf}. In the unitary gauge, which as noted above, is assumed
when passing from the Abelian Higgs to the Proca action by taking $\lambda \rightarrow \infty$,
$X=A'^2$. Applying the condition on $X$ for the minimum of $V_{\rm eff}$ to
move to the symmetry restoration point in this gauge with 
$\rho_{A'}\sim m_{A'}^2 A'^2$ again gives $\rho_{A'}\sim \lambda v^4$.
This simple argument ignores the spatial/temporal dependence of the vector field, 
but also suggests that we will expect some strong backreaction on the scalar field
approaching this threshold.

In fact, when the superheating threshold is exceeded, a huge number of vortices, that
is, cosmic strings, form all at once in the dark photon magnetic field, and
most of the energy stored in the background magnetic field turns into energy of
the cosmic strings until the electromagnetic field drops well below $B_{\rm sh}'$.
These cosmic strings radiate away their energy in the form of gravitational
waves and boosted dark photons. At formation, the cosmic strings can have
energy densities that are $\mathcal{O}(\lambda/g_D^2)$ larger than
the scaling energy density, and therefore, can contribute a burst of
gravitational waves at the time of the superheated vortex forming phase
transition.

Vortices can also form in a dark photon cloud that grows around a spinning
black hole through the superradiant instability, as recently demonstrated
in~\cite{East:2022ppo}. In black hole superradiance, a dark photon with Compton
wavelength comparable to a spinning black hole will spontaneously form a
gravitationally bound cloud that grows at the expense of the black hole's
rotational energy~\cite{Pani:2012vp,Endlich:2016jgc,Baryakhtar:2017ngi,East:2017mrj,Cardoso:2018tly}. In the Proca limit, where the dark photon only has
gravitational interactions, the growth of the cloud will continue until the
black hole is spun down below the superradiant regime~\cite{brito_review,East:2017ovw,East:2018glu}. In~\cite{East:2022ppo},
it was shown, using simulations with modestly large $\lambda/g_D^2 \leq 50$,
that approaching the superheating threshold, vortex strings are formed and
subsequently accelerated by the dark electric field.  They oscillate before
being absorbed by the black hole, emitting a significant portion of the dark
photon energy density from the cloud in a ``stringy" bosenova\footnote{We note
that boss nova is usually performed on a nylon string guitar.} event
(analogous to the bosenova scenario proposed in axion superradiance~\cite{Arvanitaki:2010sy}).
Though in those cases, only a handful of
strings formed, here we discuss how the dynamics may be different when
$\lambda/g_D^2$ is very large, with up to $10^{38}$ strings forming at the same
time in the superradiance cloud, and frequent string-string interactions leading
to the exchange of energy, linear momentum, and angular momentum.  In
particular, this would allow for some strings to become gravitationally
unbound from the black hole, with potentially observable consequences. 

The paper is organized as follows. In section~\ref{sec:Abrikosov}, we review
the knowledge of these vortex solutions in the context of type II
superconductor where they were first discovered. In sections~\ref{sec:darkA} and
\ref{sec:sim_results}, we discuss, analytically and numerically, how
these vortices nucleate in dark photon dark matter from different production
mechanisms. In section~\ref{sec:evo}, we show the evolution of the vortices
after its formation and the subsequent energy dissipation.
In section~\ref{sec:superrad_cloud}, we discuss how vortices form and evolve in
a dark photon superradiance cloud. We discuss the phenomenological consequence of
these vortices both in the case of dark photon dark matter, as well as dark
photon superradiance clouds in section~\ref{sec:pheno}.  We make some
final remarks in section~\ref{sec:remarks} regarding how the dynamics studied in this
paper can occur in other physical systems.

\section{Dynamics of the Landau-Ginzburg model and the Abrikosov lattice}\label{sec:Abrikosov}

In this section, we review the dynamics of the Landau-Ginzburg model, the
Abrikosov vortex and lattice solutions of type II superconductors, and the
various phase transitions that occur as the strength of the magnetic field
exceeds some threshold~\cite{Abrikosov:1956sx,tinkham2004introduction}. We will
use the language of the Abelian Higgs model, and point out several key
differences between dark photon dark matter and type II superconductors.

\subsection{Ground state}

The Landau-Ginzburg model has a free energy that resembles that of equation
\ref{eq:abelianhiggs}, where the inverse of the mass of the dark photon
$1/m_{A'}$ is known as the London penetration depth $\delta$, and the inverse of
the mass of the radial mode is $1/m_{\rho}$ is known as the coherence length
$\xi$.\footnote{The Landau-Ginzburg model is a time-independent model of
collective excitation of non-relativistic electrons. In order to make the
comparisons between our system and superconductors most apparent, analogous to
the $\hbar = c = k_B = 1$ units used in high energy physics, in this section, we
describe the phenomenology of superconductors in $\hbar = v_F = k_B = 1$ units,
where $v_F$ is the Fermi velocity. The different energy, length, and mass scales
are related to each other with powers of the Fermi velocity.} The ratio between
the two scales $\kappa \equiv \delta/\xi = \sqrt{\lambda/g_D^2}$ determines if the
superconductor is type I or type II. For $\kappa > \frac{1}{\sqrt{2}}$, the
superconductor is type II. The Stueckelberg limit corresponds to the case of
$\kappa \rightarrow \infty$. In a type II superconductor, there are two
critical fields. The critical field (the corresponding equation in the case of
a dark photon $U(1)'$ is shown in parentheses, also see table~\ref{tab:variabledefinitions} in appendix~\ref{sec:notation} for a complete list of the terminology relevant to both systems)
\begin{equation}
H_{\rm c2} = \frac{\Phi_0}{2 \pi \xi^2} \quad  \left( B_{\rm c2}' = \frac{\lambda v^2}{g_D}    \right)
\end{equation}
is the field strength at which superconductivity is completely lost, that is,
the $U(1)$ symmetry is restored globally. Here $\Phi_0 = \pi/e$ is the magnetic
flux quantum. At this field strength, the mass of $\Phi$ (near $\Phi= 0$)
\begin{equation}
    m_\Phi^2 = -\lambda v^2 + g H
\label{eq:massterm}
\end{equation}
goes to zero and $\langle |\Phi|\rangle = v(1-g H/\lambda v^2)^{1/2} \rightarrow 0
$ becomes the global minimum, where $g=2e$ is the coupling of the Cooper pair to the photon. However, the field strength $H_{c2}$ is not the
scale below which the superconducting state is the ground state. A state with a
lattice of vortices (the Abrikosov lattice) has a lower free energy compared to the
state where all the magnetic field lines are expelled (the Meissner effect) when
the average magnetic field strength is larger than 
\begin{equation}
    H_{c1} \approx \frac{\Phi_0}{4 \pi\delta^2} \log \kappa \quad  \left(B_{\rm c1}' = g_D v^2 \log \frac{\lambda}{g_D^2}\right) .
\end{equation}
In the limit of large $\kappa$, this field strength is $\mathcal{O}(\log \kappa/\kappa^2)$ smaller than $H_{c2}$.

\begin{figure}
    \centering
    \includegraphics[width=0.5\textwidth]{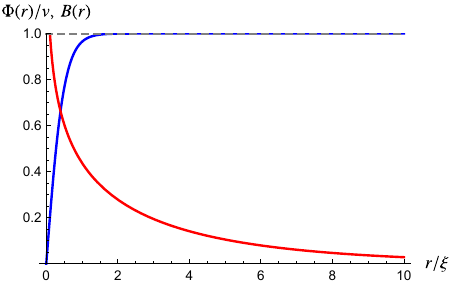}
    \caption{The field profile of an isolated vortex. The blue line is the
    profile of the radial mode $|\Phi(r)|$, which has a characteristic extent 
    $\xi$, while the red line is the profile of the magnetic field, which has a
    characteristic extent $\delta$. Here we take $\delta/\xi = 10$.}
    \label{fig:Profile}
\end{figure}

At field strengths $H \gtrsim H_{c1}$, the ground state is comprised of a lattice of vortices. Each of the vortices has the following field profile in isolation in the $\kappa \rightarrow \infty$ limit
\begin{equation}
    H(r) = \frac{\Phi_0}{2\pi \delta^2} K_0\left(\frac{r}{\delta}\right),
\end{equation}
where $K_0$ is the zeroth-order Hankel function of imaginary argument. This function has the following asymptotic behaviour (see figure~\ref{fig:Profile})
\begin{align}
    H(r) \approx \frac{\Phi_0}{2\pi \delta^2}\left(\frac{\pi \delta}{2 r}\right)^{1/2} e^{-r/\delta}, \quad r\gg \delta \nonumber\\
    H(r) \approx \frac{\Phi_0}{2\pi \delta^2} \left(\log \frac{\delta}{r}+0.12\right), \quad \xi \ll r \ll \delta.
\end{align}
The first equation above suggests that the size of the vortex is the magnetic
penetration depth (the inverse of dark photon mass). The $\log
(\delta/r)$ term in the second equation gives rise to a logarithmically divergent energy per unit
length stored in the magnetic field in the limit where the string core size is
taken to be zero, in agreement with the logarithmically large string tension in
section~\ref{sec:intro}. 

\begin{figure}
    \centering
    \includegraphics[width=0.5\textwidth]{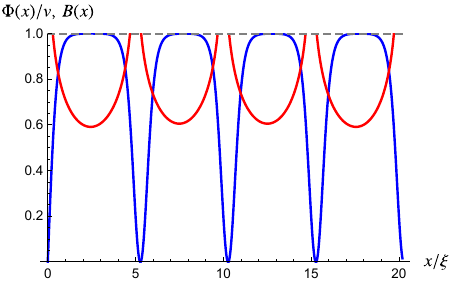}
    \caption{The field profile of a vortex lattice in a 1D slice. The blue line
    is the profile of the radial mode $|\Phi(x)|$, which has a characteristic
    extent of $\xi$, while the red line is the profile of the magnetic field,
    whose periodicity is the lattice spacing $a_L$. Here we take $\delta/\xi = 10$ and $a_L/\xi
    =5$.}
    \label{fig:Profile2}
\end{figure}

At field strengths of $H_{c1}\ll H \ll H_{c2}$, the ground state is known as the Abrikosov lattice, where the vortices line up in a 2D lattice with lattice spacing 
\begin{equation}
    a_{L} = c_L \left(\Phi_0/H\right)^{1/2} \ll \delta ,
\end{equation}\label{eq:latticespacing}
where $c_L$ is an $\mathcal{O}(1)$ number which depends on the structure of the
lattice of the vortices (see figure~\ref{fig:Profile2})\footnote{We have
neglected the differences between $B$ and $H$, which is a good approximation in
the limit $H_{c1}\ll H \ll H_{c2}$.}. The force between two vortex lines is 
repulsive, due to the overlap of the magnetic field when $a_L \ll \delta$
($\kappa >1/\sqrt{2}$, to be exact). However, when $a_L \ll \delta$, a large
number of vortices overlap with each other, and the system exhibits complicated
vortex matter dynamics (
see~\cite{RevModPhys.90.015009}).

\subsection{Superheating}

\begin{figure}
    \centering
    \includegraphics[width=0.5\textwidth]{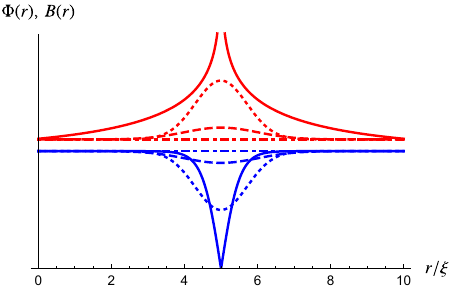}
    \caption{The field profile evolution when the field strength approaches the
    superheating field. The magnetic field $B$ (red lines) and the scalar field magnitude
    $|\Phi|$ (blue lines) are both constant in the beginning (dot-dashed). A
    small perturbation in the magnetic field (dashed) leads to a localized
    $\Phi$ profile. This causes magnetic field to further cluster in the region
    where $\langle\Phi\rangle$ is small (dotted) and the scalar profile to
    further develop a dip. The magnetic field grows logarithmically to infinity
    in this localized region, eventually pushing $\Phi$ to zero, creating a
    vortex. The solid lines show the profiles of a localized vortex as
    reference.}
    \label{fig:Insta}
\end{figure}

The phase transition between the vortex lattice phase and the Meissner phase is
a first order phase transition when $H \gtrsim H_{c1}$. This is to be expected
since the two phases have very different spacetime symmetries, as well as
topologies. In particular, when $\kappa\rightarrow \infty$, the vortex solution
has a profile with typical size $\xi$, while the profile
for the vector potential has a typical size $\delta$. The system can remain in
the Meissner phase until $H= H_{\rm sh}$, the superheating field, when this
phase transition becomes a second order phase
transition~\cite{Galaiko1966FormationOV,osti_4551407,liarte2016ginzburg,PhysRev.170.475,PhysRevB.65.144529}.
The superheating field is
\begin{equation}
H_{\rm sh} = C_{\rm sh} \Phi_0/\delta\xi \quad  \left( B_{\rm sh}' = \lambda^{1/2} v^2\right) ,    
\end{equation}
where $C_{\rm sh}$ is an $\mathcal{O}(1)$ coefficient that depends on the
geometry of the system. At the superheating field, a linear combination of the
vector potential and the order parameter becomes tachyonic, and the phase
transition becomes second order. A qualitative picture of this instability is
shown in figure~\ref{fig:Insta}. As the magnetic field approaches $H_{\rm sh}$,
a localized perturbation of magnetic field causes the $|\Phi|$ field to also
develop a localized profile (see equation~\ref{eq:massterm}). This localized
dip in $\langle |\Phi|\rangle$ causes the magnetic field to further cluster
(similarly to the Meissner effect), leading to a runaway effect, and a vortex
is formed despite the initial background magnetic field being much smaller than
the critical field $H_{c2}$. In this instability, the clustering magnetic field
only needs to overcome the mildly growing gradient energy in the fields,
leading to the formation of a defect in a field that seems decoupled in the
$\lambda/g^2 \rightarrow \infty$ limit.

Recent studies~\cite{PhysRevB.83.094505} have identified the most tachyonic mode as
a vortex like excitation with a typical momentum $k_c \simeq
\kappa^{3/4}/\delta$, instead of either $1/\delta$ or $1/\xi$. Compared to the
lattice spacing $a_L$ at the superheating field computed from equation
\ref{eq:latticespacing},
\begin{equation}
    k_c a_L \sim \kappa^{1/4}.
\end{equation}
This further highlights that the vortices nucleated at the superheating field
have profiles and structures that are very far from the eventual equilibrium
state. As we will see, this can have distinct phenomenological consequences for
dark photon dark matter. 

At field strengths below the superheating field, the phase transition is first
order. The barrier in free energy between the two phases was first studied
in~\cite{Galaiko1966FormationOV}, where it was found that the barrier in free
energy is $\Delta F \sim \Phi_0^2 \xi(\log \kappa)^2/\delta^2\sim v
(\log \kappa)^2/\lambda^{1/2}$ for a thermal phase transition. The vacuum phase
transition, which is more important for our understanding of the stability
of dark photon dark matter in the late universe, has not, as far as the authors know,
been well understood analytically due to the very limited symmetry of
the bounce action. 

\subsection{Melting phase transition}\label{sec:melting}

Following a superheated phase transition, vortex lattices in type II
superconductors exhibit a melting phase transition for large ${\lambda/g_D^2}$.
The melting phase transition in the context of a vortex lattice is the
formation of many small vortex loops such that the vortex-vortex correlation
functions decays as the vortex separation grows, and as a result, the
translational symmetry broken by the background magnetic field and the vortex
lattice is
restored~\cite{zeldov1995thermodynamic,james2021emergence,liu1991kinetics,PhysRevB.8.3423}. 

The superheated phase transition occurs between two states that have very
different free energy density, which suggests that the superheated phase
transition produces a network of vortices that is very far from the ground
state configuration in the same background magnetic field. After the phase
transition, a network of vortex lines with total length that is
$\mathcal{O}(\kappa)$ times larger than the length of vortex lines of the
ground state (in the same background magnetic field) can be formed. This
network of vortices, through vortex reconnections, produces small loops of
vortices with random locations and orientations. 
As the melting phase transition eventually completes, 
the mostly straight lines that align with the external magnetic field 
have lengths that are $\mathcal{O}(1/\kappa)$ of the total
length of vortex lines in the system. The small loops therefore dominate the
vortex-vortex correlation functions, and an approximate translational symmetry
is restored in this vortex fluid despite the background magnetic field.

\section{Vortex formation in dark photon dark matter}\label{sec:darkA}
The existence of these vortex solutions and vortex formation mechanisms raises
concerns about whether the light $U(1)'$ dark photon has a viable production mechanism to
be a dark matter candidate. The best-motivated production mechanism for light
dark photon is inflationary production~\cite{Graham:2015rva}, which produces a
dark photon longitudinal mode through inflationary perturbations. Recently,
there have been proposals for a new tachyonic particle production
mechanism~\cite{Agrawal:2018vin}, which predominantly generates magnetic field
that later redshifts to be cold dark matter. In this section, we show how
vortices form in both cases. 

\subsection{Vortex Formation: Longitudinal mode}

In the case of inflationary production, vortex formation happens as long as the
inflationary scale $H_I > v$. In the limit of infinite $\lambda$, the effective
action is a combination of the Proca action plus the Nambu-Goto string action.
The production rate of Nambu-Goto string during
inflation~\cite{PhysRevD.44.340}
\begin{equation}
    \Gamma \sim \exp [-4\pi \mu/H_I^2],
\end{equation}
where $\mu \sim \pi v^2 \log (\min[\lambda/g_D^2,m_{\rho}/H_I])$ is the string
tension. The logarithm comes from the logarithmically divergent string tension, where the UV
scale is the radius of the string core (scalar profile), while the IR scale is
the larger of the mass of the dark photon, or the Hubble scale during inflation.
For $v \lesssim H_I$, there will be $\mathcal{O}(1)$ strings produced in each
Hubble volume per Hubble time. Even if $v \gtrsim H_I$, as we elaborate in
appendix~\ref{sec:NonMCoupling}, a string network that
eventually approaches a scaling solution can still be produced.

In the case of longitudinal mode production, the formation does not suffer from
a suppression due to superheating because the longitudinal mode fluctuation
itself generates the vortex. The $\Pi$ field has a $2\pi$-periodicity of $\Pi/v
\rightarrow \Pi/v+ 2 \pi$. The longitudinal mode $A_L$ inherits this periodicity. A
large amplitude of $\Pi$ and $A_L$ means that the field winds around many
times, which is equivalent to producing a large number of vortices. In order to
not produce these vortices, we need the scalar VEV $v > H_I$. Given that the
dark matter density is~\cite{Graham:2015rva}
\begin{equation}
     \Omega_{\rm A'} = \Omega_{\rm DM} \times \sqrt{\frac{m_{A'}}{6\times 10^{-6} \,{\rm eV}}}\left(\frac{H_I}{10^{14} \,{\rm GeV}}\right)^2,
\label{eq:graham}
\end{equation}
we have
\begin{equation}
  g_D = \frac{m_{A'}}{v} \leq \frac{m_{A'}}{H_I} = 2\times 10^{-22} \left(\frac{m_{A'}}{\rm eV}\right)^{5/4} .
\label{eq:requirement}
\end{equation}
For a kinetic mixing with the standard model photon of $\epsilon \sim g_De/(16 \pi^2)$~\cite{Holdom:1985ag}, this
corresponds to $\epsilon < 10^{-24}$, about ten orders of magnitude beyond
current experimental capabilities.

One possible way to evade this limit is to introduce a charge hierarchy of
$10^{10}$ between the particle that runs in the loop to generate the kinetic
mixing and the Higgs~\cite{Gherghetta:2019coi}. These models are constructed
mainly to explain the smallness of the kinetic mixing parameter as compared to
the gauge coupling $g_D$. To evade our constraints, one needs to run the
clockwork mechanism backwards.  We makes some additional remark on this possibility in
appendix~\ref{sec:clockwork}. A second possible way out is to introduce a large
non-minimal coupling to gravity that suppresses string production. However, this does
not solve the problem of producing enough dark photon dark matter; we elaborate on this
possibility in appendix~\ref{sec:NonMCoupling}.

The quartic $\lambda\rightarrow \infty$ limit is the simplest way to achieve a
Stueckelberg mass for a $U(1)'$ dark photon, which works perfectly in scenarios
where collective effects are inaccessible.  A few other UV completions
have been studied to generate a small dark photon mass and kinetic
mixing~\cite{Goodsell:2009xc}, motivated by string theory. In these cases, the
lightness of the radial mode (or other charged particles) is not related to the
smallness of the dark photon mass, the dark gauge coupling, and the kinetic
mixing in the same way as the cases studied in this paper. However, there
generically exist new states below the Hubble scale in
eq.~\ref{eq:graham}, whose dynamics can significantly disrupt the dark
photons, either by vortex formation, in the cases where there is a charged
scalar, or as in the scenarios studied in~\cite{Arvanitaki:2021qlj}, if there are only
charged fermions. We will not comment further on these scenarios since there are
no concrete models, as far as the authors know, where a small dark photon
Stueckelberg mass is generated in a theory where the string scale is above the
Hubble scale during inflation implied in equation~\ref{eq:graham}. The weak
gravity conjecture~\cite{Arkani-Hamed:2006emk,Reece:2018zvv} places some doubt
on whether such a scenario can be reliably constructed in string theory to
produce a light dark photon as dark matter. Combining the weak gravity conjecture $g_D \gtrsim
m_{\rho}/m_{\rm Pl} \gtrsim H_I/m_{\rm Pl}$ and $g_D \leq m_{A'}/H_I$
with eq.~\ref{eq:graham} suggests that $m_{A'} \gtrsim 50\,{\rm GeV}$.

\subsection{Vortex Formation: Transverse mode}
In the background of a transverse dark photon, the dynamics of vortex formation
is similar to vortex formation in a superconductor. The presence of the magnetic field (flux) is essential. However, the production of vortices
is not significantly hindered by the presence of a similar-sized or larger (on
average) electric field, as long as there is no Lorentz transformation that can
remove the magnetic field everywhere, and, in fact, the strength of the electric
field can aid in crossing the superheating threshold. This is demonstrated by 
the case of the superradiant dark photon cloud, where the vector field is
electrically dominated, and vortex formation happens when $-F'^2$ approaches
$\sim B'^2_{\rm sh}$. We will describe this in more detail in
section~\ref{sec:superrad_cloud}.

\paragraph{Magnetic mode} In a background magnetic field, vortices should
form when the background magnetic field exceeds a similar superheating field of
$B'_{\rm sh}$, just like in the case of a superconductor. However, there are
some key differences between vortex formation in superconductors and in dark
photon dark matter. The most important difference is that in the case of
superconductor, the phase transition happens between the Meissner phase and the
Abrikosov lattice phase, where there is originally zero magnetic field inside
the bulk of the superconductor (the Higgs phase). All vortices enter from the edge
of the superconductor. However, in the case of dark photon dark matter, there
is already an oscillating electric field and magnetic field inside the bulk
where the dark photon is in the Higgs phase. The second main difference is that
a superconductor always has an edge where vortex lines can end on, since beyond
the edge of the superconductor, the photon is massless. However, in the case of
dark photon dark matter, the whole Universe is in the Higgs phase and the
vortices have to form as string loops. Lastly, in both dark photon dark
matter and a dark photon superradiance cloud, there are usually electric fields
of similar size or larger in the same system, which can affect vortex
formation. Due to the above-mentioned differences, in the next section 
we perform several numerical studies of vortex formation in dark photon dark matter. 

\paragraph{Electric mode}

It is unclear what the production mechanism for a transverse mode of a dark photon that is purely electric
field would be, since this would mean such modes can
be produced with zero momentum dependence\footnote{
In the example of the dark photon superradiance cloud, vortices form in the
direction of the magnetic field despite the presence of a stronger electric
field~\cite{East:2022ppo}.}. 
On the other hand, if there were some unknown mechanism for producing
transverse mode dark photon dark matter at the time of CMB, then the
dark photon dark matter would be predominantly electric field mode due to
redshifting. 

In a purely curl-free electric field background, vortex formation does not
happen. As the field strength of the electric field reaches $g_D E' \gtrsim m_{A'}
m_{\rho}$ (or equivalently, $E' \gtrsim B'_{\rm sh}$), the heavier scalar field and the lighter dark photon field
oscillators become strongly coupled and energy can be transferred between the
dark photon and the radial mode. When the electric field is small, the energy that
oscillates into the radial mode grows as $E'^4$ when $E'^2 < \lambda v^4$. As
the dark photon field continues to grow, the system becomes non-linear.
However, since both energies redshift like matter, it is unclear if these
oscillations significantly change the dark photon density after $E'^2$ eventually drops below $\lambda v^4$.

\section{Vortex formation in magnetic dark photon production}\label{sec:sim_results}

In this section, we study numerically how vortices form in dark photon dark matter in production mechanisms where the dark photon is produced initially as dark magnetic field. 
As shown in some recent studies~\cite{Agrawal:2018vin,Co:2018lka,Dror:2018pdh,Bastero-Gil:2018uel}, dark photon dark matter can also be
produced in the late Universe by invoking a coupling between the dark photon and rolling 
axion field $a$ of the form 
\begin{equation}
    \mathcal{S} = \int {\rm d}^4 x \left(\ldots + \frac{a}{f} F'_{\mu \nu}\tilde{F}'^{\mu \nu}\right) \ ,
    \label{eqn:axion_coup}
\end{equation}
where $f$ is the axion decay constant. In the background of a time dependent axion 
field $\dot{a}/f \neq 0$, the dark photon dispersion relation is 
\begin{equation}
    \omega^2 = (|\vect{k}|\pm \dot{a}/2f )^2 - (\dot{a}/2f )^2 + m_{A'}^2 -H^2/4.
\end{equation}
Thus, modes with wavenumbers in the range $k\in [k_-,k_+]$ where
\begin{equation}
k_{\pm} = \dot{a}/2f \pm \sqrt{(\dot{a}/2f )^2-m_{A'}^2+H^2/4}
\end{equation}
can grow exponentially in time, with the fastest growing mode having wavenumber
$k_p=(k_++k_-)/2$.  This instability can efficiently transfer energy from the
axion field $a$ to the dark photon electric and magnetic fields if $\dot{a}/2f
\geq m_{A'}, H$. The exact fraction of energy in the electric and magnetic
field at any time depends on the relative size of the parameter $\dot{a}/2f$,
$m_{A'}$ and $H$, and the shape of the axion potential. Nevertheless, at the
time when the tachyonic instability stops and the $A'$ density is at its peak,
$\omega \sim k \geq m_{A'}$ and a significant portion of the energy is stored
in the magnetic field. A large dark magnetic field (in particular,
$\vect{E}'\cdot \vect{B}'$) can backreact on the rolling axion field. The
leading effect of this is to slow down the rolling axion (decreasing $\dot{a}$), which
stops dark photon production. Throughout the paper, we restrict to regions of the parameter
space where the backreaction of the gauge field on the rolling axion field
is small.  

As suggested in section~\ref{sec:Abrikosov}, vortex formation can occur if the
magnetic field generated by the tachyonic instability grows sufficiently large.
We use numerical simulations to explore the formation and subsequent dynamics
of the vortices.  Since this is our primary interest, we will ignore the
cosmological evolution and backreaction of the axion field, and numerically
evolve the Abelian-Higgs equations in the Lorenz gauge ($\partial_\mu
A'^{\mu}=0$) with a source term due to a homogeneous rolling axion background.
The vortex formation dynamics is independent of the reshifting of the gauge
fields, while the subsequent evolution of the strings post-vortex formation
will depend on cosmic expansion. Given the large amount of energy that needs
to be dissipated to reach close to a scaling solution for moderate
$\lambda/g_D^2$, it is likely very challenging to track the full dynamics in an
expanding universe based on what has been achieved in the literature for
similar and slightly simpler systems~\cite{Buschmann:2021sdq}. See
section~\ref{sec:num} for details on the evolution equations and numerical
methods. 

\subsection{Two-dimensional results}
We begin with results assuming a translational symmetry in one direction.  This
makes tackling the large $\lambda/g_D^2$ regime more computationally tractable,
and more straightforwardly connects to the Abrikosov lattice picture of
section~\ref{sec:Abrikosov}. However, we also perform fully 3D simulations, to
highlight the similarities and differences from the 2D case, which we discuss
in the next section.  

\begin{figure}
\begin{center}
\includegraphics[width=0.5\columnwidth,draft=false]{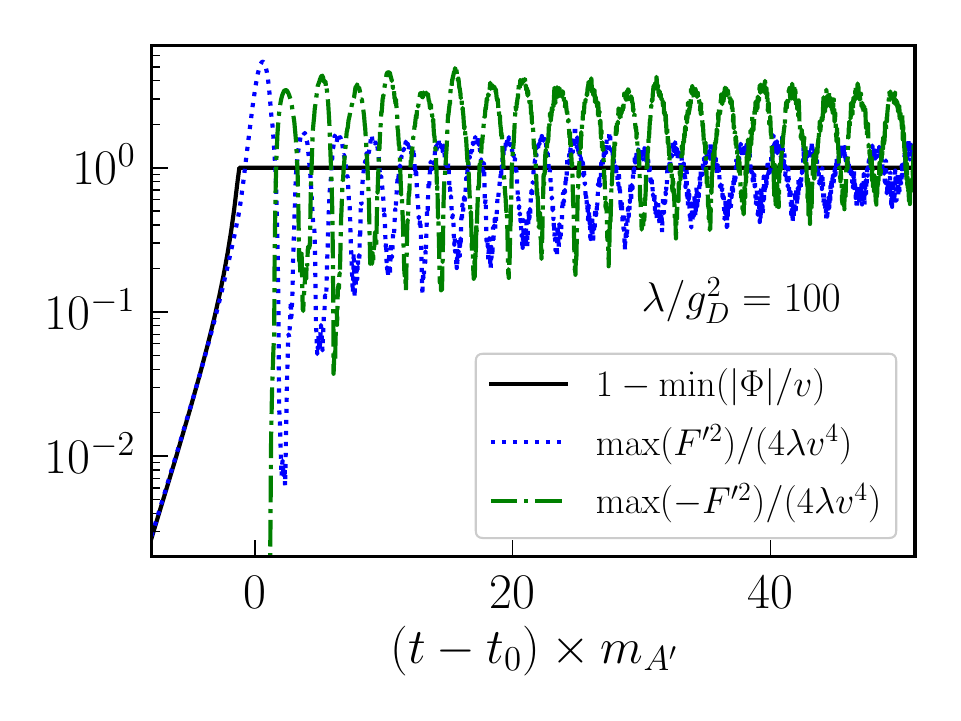}
\end{center}
\caption{
    Maximum and minimum field values as a function of time.  We show the
displacement from the VEV of the minimum scalar field magnitude (solid black
line), which tracks the maximum of the field tensor contracted with itself
$F'^2$ (dotted blue line) closely during the linear instability phase.  We also
show the maximum values of $-F'^2$ (dashed-dotted green line)  when this
quantity is positive.  Here $F'^2\equiv F'^{\mu \nu} F'_{\mu
\nu}=2(B'^2-E'^2)$, so $F'^2>0$ corresponds to magnetic dominance, and $F'^2<0$
corresponds to electric dominance.
\label{fig:field_comp}
}
\end{figure}

As the vector field grows exponentially due to the tachyonic instability, the
scalar field becomes increasingly displaced from its VEV towards lower
magnitudes in the regions of high dark photon field.  We illustrate a
representative case with $\lambda/g_D^2=100$ in figure~\ref{fig:field_comp}.
There, the maximum of $1-|\Phi|/v$ can be seen to track the maximum of
$F'^2/(4\lambda v^4)$. This agrees with the shift in the minimum of $V_{\rm eff}$
in eq.~\ref{eqn:veff} since $F'^2\approx m_{A'}^2 X$ in the linear instability phase. 
This persists until $\Phi=0$ somewhere in the domain, which marks the
onset of vortex formation. Shortly after vortex formation, we switch off the
axion source term (see section~\ref{sec:num} for details), which halts the growth
of the vector field. In the exponential growth phase, the vector field is
everywhere magnetically dominated, $F'^2>0$. However, post-vortex formation,
the vector field decays as energy is transferred from the vector to scalar
sectors. One can also see oscillations occurring at a frequency of $\sim
2m_{A'}$ between regions of strong electric dominance (i.e. $F'^2$ negative
and large in magnitude) and magnetic dominance.  We will describe this in more
detail below. 

\begin{figure}
\begin{center}
\includegraphics[width=0.39\columnwidth,draft=false]{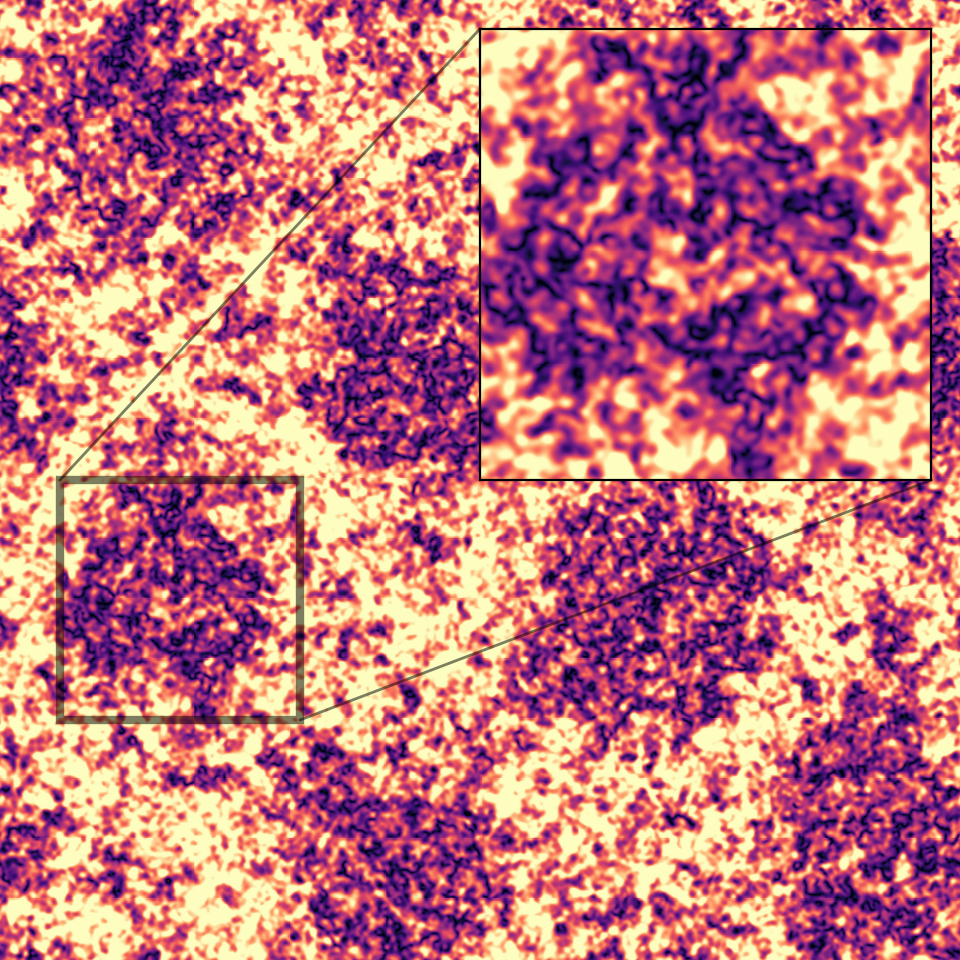}
\raisebox{0.7cm}{\includegraphics[draft=false,trim=200 0 0 0, clip,width=0.1\columnwidth,]{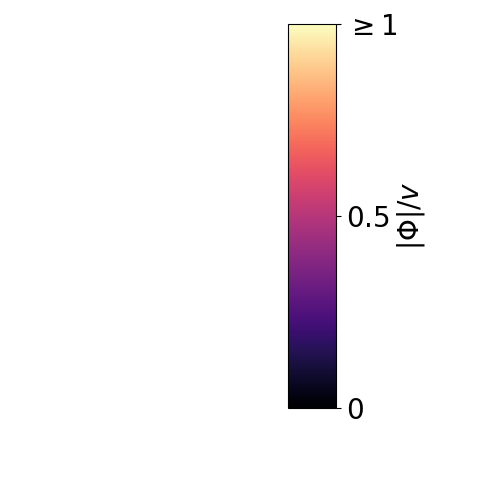}}
\includegraphics[width=0.39\columnwidth,draft=false]{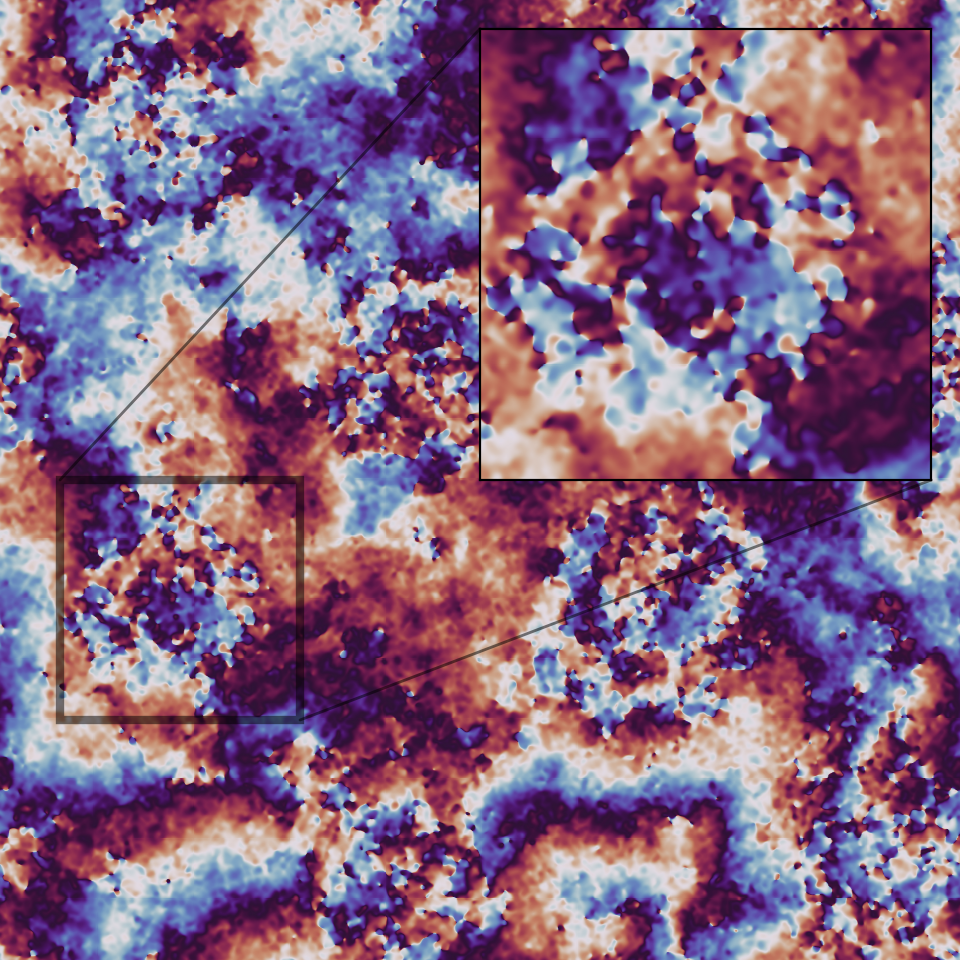}
\raisebox{0.7cm}{\includegraphics[width=0.1\columnwidth,draft=false,trim=200 0 0 0, clip]{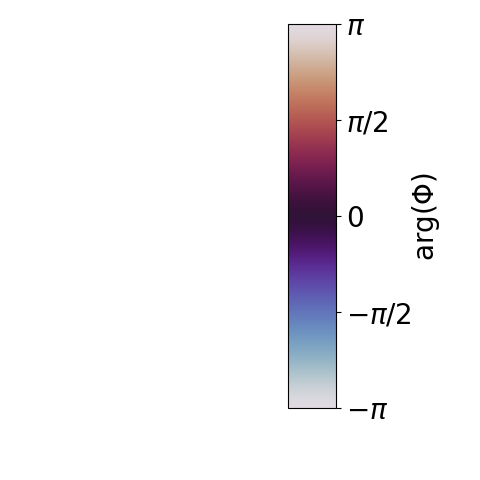}}
\end{center}
\caption{
Snapshot at late times of the complex scalar $\Phi$ from the same case with $\lambda/g_D^2=100$ shown in figure~\ref{fig:field_comp}.
The left panel shows the magnitude of $\Phi$, while the right panel shows the complex phase. In both panels, the inset at the top right shows
 a zoom-in of a region where a number of vortices have formed.
The length of the domain is twice the wavelength of the dominant mode of the axion instability, 
the latter of which is approximately equal to the Compton wavelength of the dark photon ($L=4\pi/k_p$ where
$k_p \approx 1.055 m_{A'}$).
A translational symmetry is assumed in the out-of-page direction.
\label{fig:phi_lam100}
}
\end{figure}

\begin{figure}
\begin{center}
\includegraphics[width=0.44\columnwidth,draft=false]{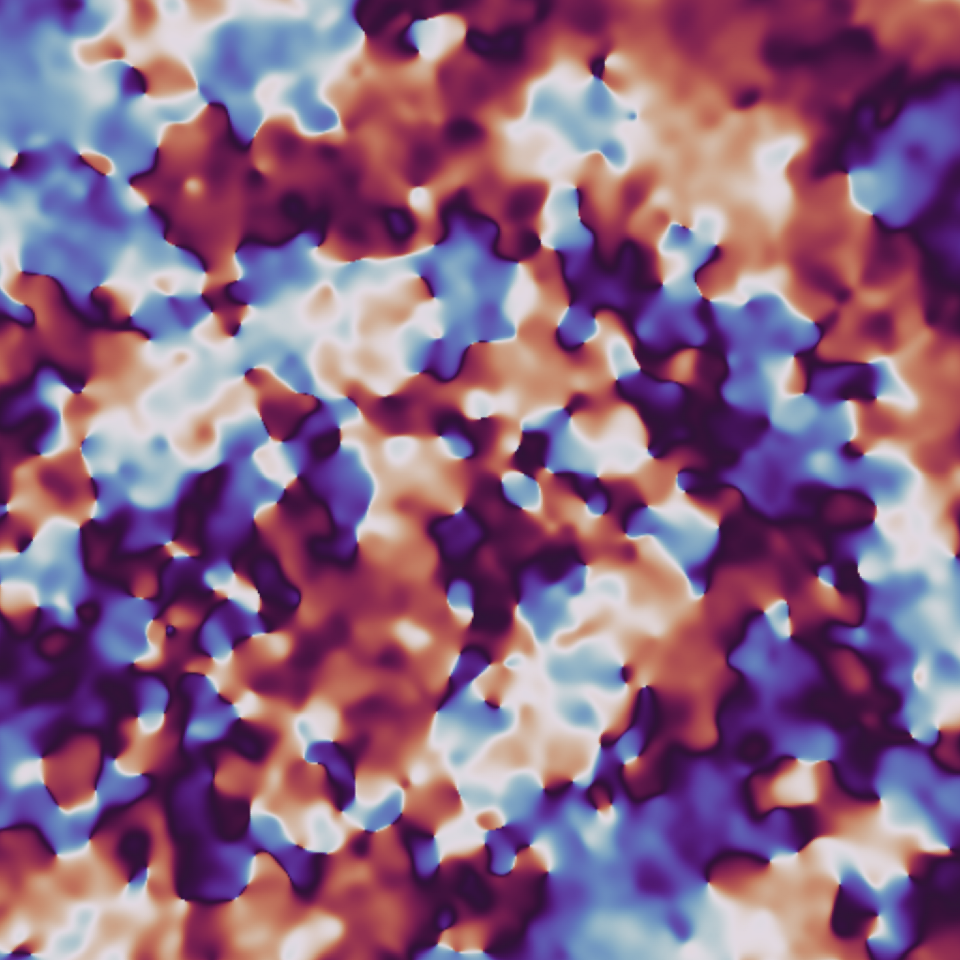}
\includegraphics[width=0.44\columnwidth,draft=false]{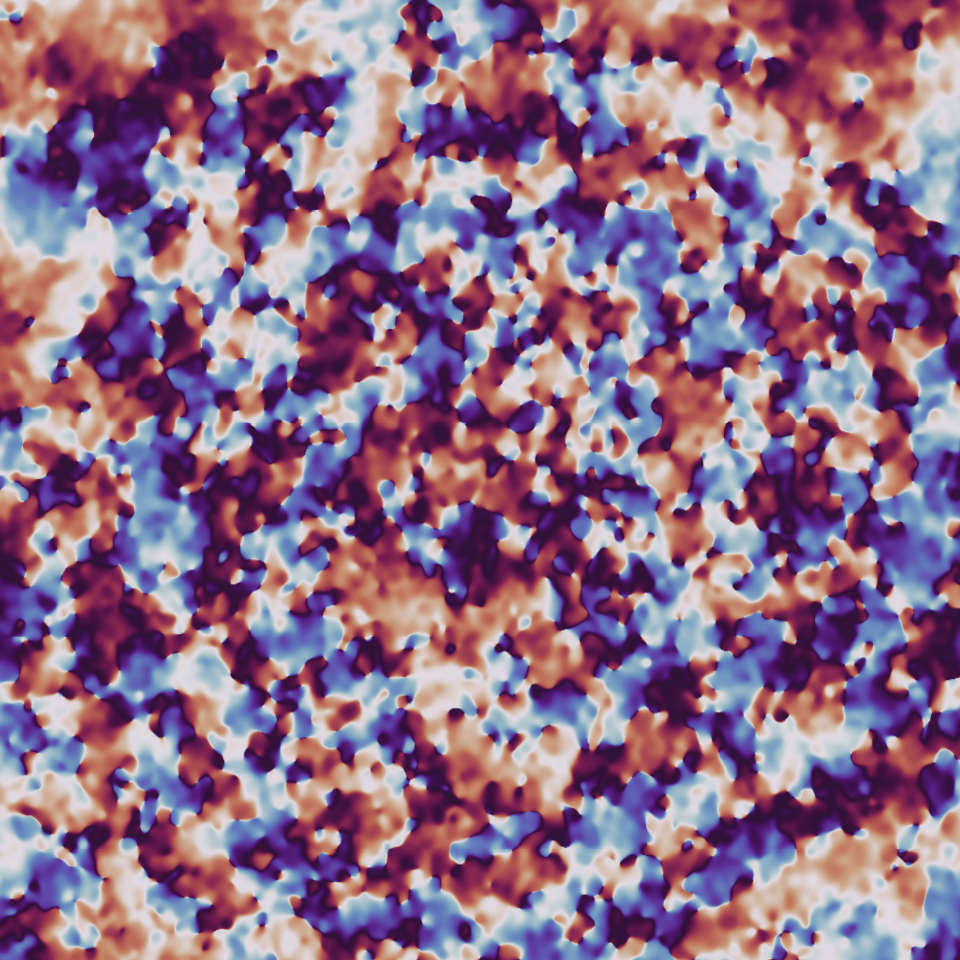}
\raisebox{0.7cm}{\includegraphics[width=0.1\columnwidth,draft=false,trim=200 0 0 0, clip]{argphi_colorbar}}
\end{center}
\caption{
\label{fig:phi_lam_comp}
Snapshot of the complex phase of $\Phi$ for $\lambda/g_D^2=100$ (left) and $\lambda/g_D^2=400$ (right) at approximately the same time 
($\approx 30/m_{A'}$ after vortices form) over a portion of the domain of length $\approx \pi/m_{A'}$.
This illustrates the increasing density of vortices as $\lambda/g_D^2$ is increased.
}
\end{figure}

The vortices are characterized by places where the magnitude of $\Phi$ goes to
zero, and the complex phase goes through some non-zero multiple of $2 \pi$ when
encircling the point.  This can be seen in figure~\ref{fig:phi_lam100}, where the
scalar field configuration is illustrated.  As $\lambda/g_D^2$ is increased,
the main difference is that the characteristic size of the vortices decreases,
and their density increases. This is illustrated in
figure~\ref{fig:phi_lam_comp}, where we compare a case with $\lambda/g_D^2=100$
to $\lambda/g_D^2=400$.  

\begin{figure}
\begin{center}
\includegraphics[width=0.49\columnwidth,draft=false]{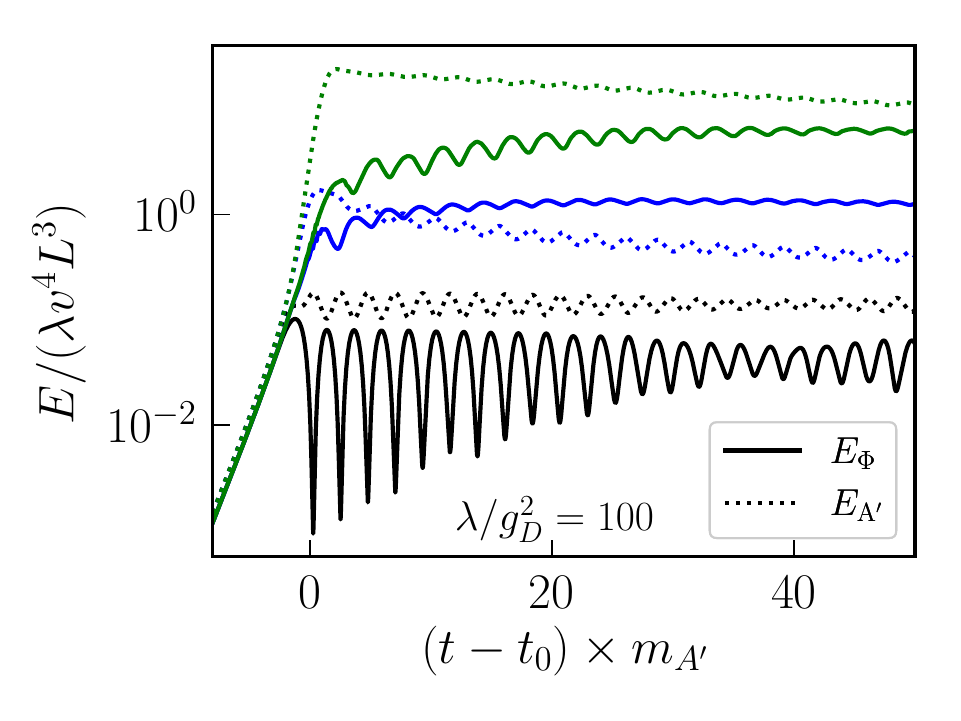}
\includegraphics[width=0.49\columnwidth,draft=false]{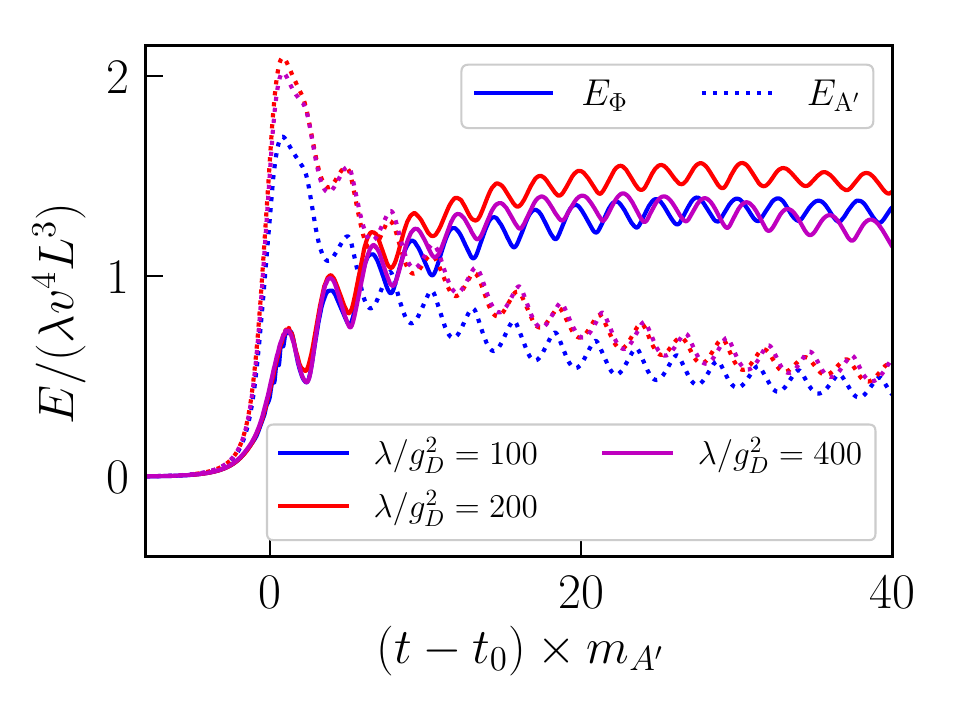}
\end{center}
\caption{
    Energy in the scalar field (including interaction energy with the vector; solid lines) and
    energy in the vector field (not including interaction energy; dashed lines) as a function of time.
    \emph{Left Panel:} We show three different cases with $\lambda/g_D^2=100$ where the time $t_0$ when the axion instability is shut off increases
    by increments of $\approx 1.2/m_{A'}$ going from black, to blue, to green.
    In the case when the axion instability is shut off the earliest (black lines), no vortices form, while 
    in the other two cases, they do.
    For the x-axis, time is scaled to the value of $t_0$ for the intermediate case (blue curves).
    \emph{Right Panel:} We compare three different cases with $\lambda/g_D=100$, 200, and 400, and with $t_0$ 
    chosen to give approximately the same value of $E_{\Phi}/\lambda$ at the time when the instability is shut off. The blue curves
    ($\lambda/g_D^2=100$) are the same case as in the intermediate case in the left panel.
\label{fig:e_tcomp}
}
\end{figure}

After the vortices form, the dark photon field does work on them,
transferring energy from the scalar to the vector sector. We discuss this dynamics analytically in more detail in section~\ref{sec:evo}.
To quantify this, we compute the energy $E$, or equivalently, 
average energy density $\langle \rho \rangle$, for the vector field
\begin{equation}
E_{A'}=\langle \rho_{A'} \rangle L^3 = \frac{1}{2}\int \left (E'^2+B'^2 \right) dV \ ,
\end{equation}
and the scalar field
\begin{equation}
    E_{\Phi}=\langle \rho_{\Phi} \rangle L^3 = \int \left [\frac{1}{2}|D_t\Phi|^2 +\frac{1}{2} |D_i \Phi|^2+ \frac{\lambda}{4}\left(|\Phi|^2-v^2\right)^2 \right] dV \ ,
\end{equation}
where we have included the interaction energy (i.e. terms involving both $A'$
and $\Phi$) in the latter.  We show these quantities as a function of time in
the left panel of figure~\ref{fig:e_tcomp} for three different cases where we
vary the time at which the axion instability shuts off, and hence the magnitude
to which the vector field grows.  When no vortices form (black curves in left
panel of figure~\ref{fig:e_tcomp}), there is little energy transfer between the
vector and scalar sectors, and the scalar energy remains subdominant. In this
case, the presence of the radial mode and the energy transfer between the two
components do not significantly affect the evolution of the energy density.
Such is also the case in a background electric field.  When vortices do form,
the vector field can do work on them, with energy moving from the vector into
the scalar sector until the latter becomes dominant (see
section~\ref{sec:electric} for further discussion of this).  In this regime,
this is largely independent of the exact value of $\lambda/g_D^2$, as is
demonstrated in the right panel of figure~\ref{fig:e_tcomp}, where we compare
several cases with $\lambda/g_D^2=100$, 200, and 400.

\begin{figure}
\begin{center}
\includegraphics[width=0.39\columnwidth,draft=false]{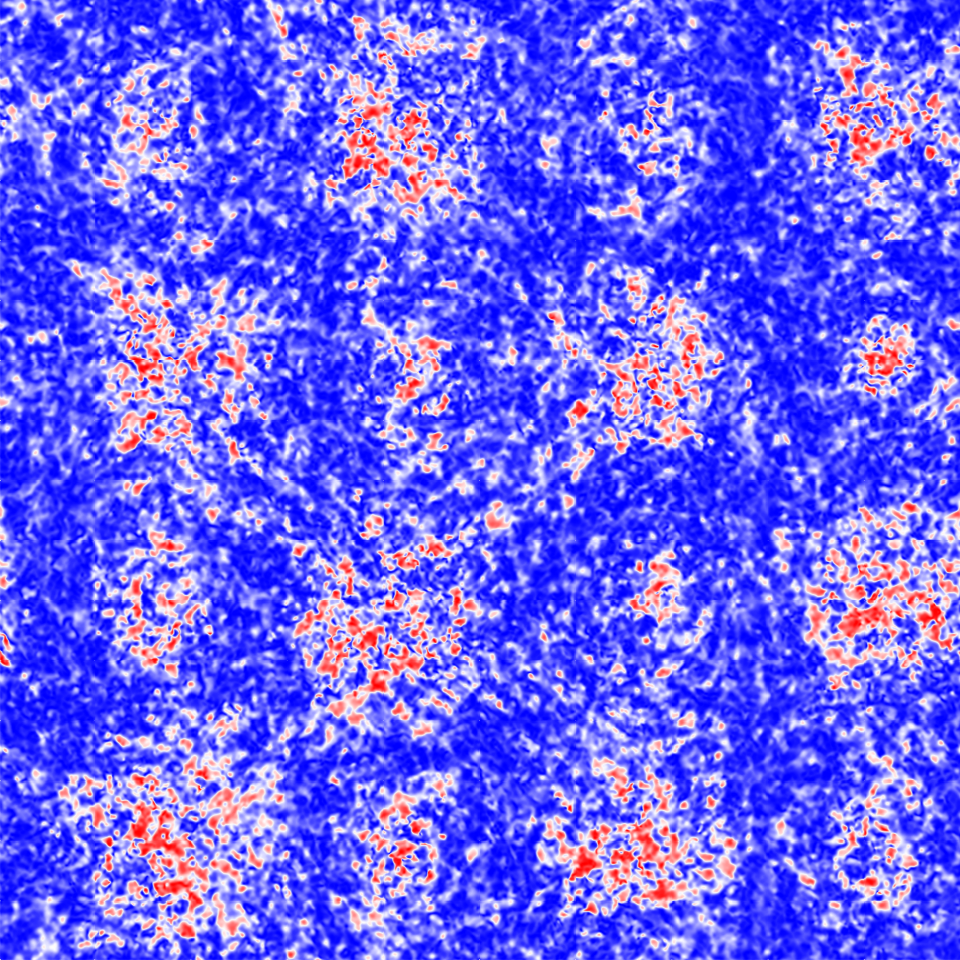}
\raisebox{0.7cm}{\includegraphics[draft=false,trim=200 0 0 0, clip,width=0.1\columnwidth,]{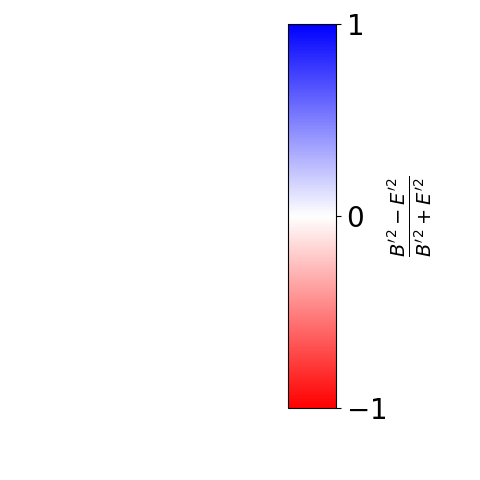}}
\includegraphics[width=0.39\columnwidth,draft=false]{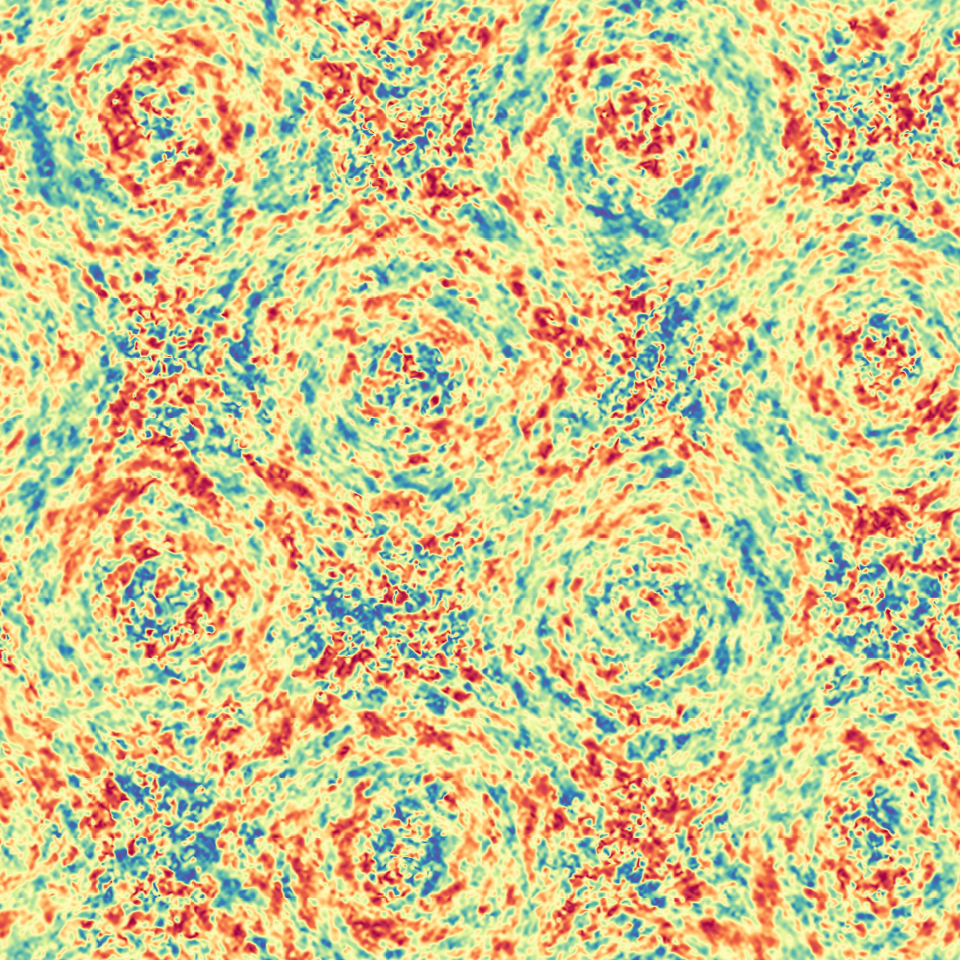}
\raisebox{0.7cm}{\includegraphics[draft=false,trim=200 0 0 0, clip,width=0.1\columnwidth,]{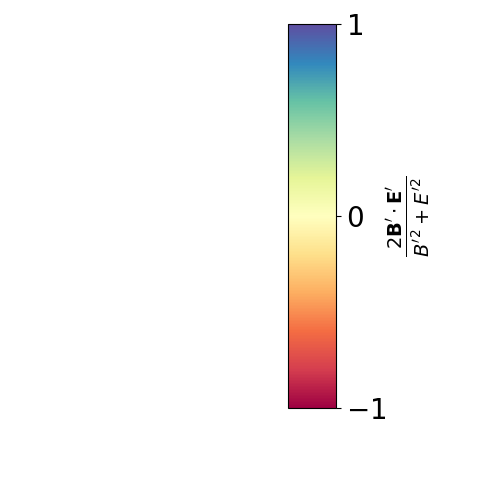}}
\end{center}
\caption{
Snapshot at late times of the vector field from the same case with $\lambda/g_D^2=100$ shown in figures~\ref{fig:field_comp} and~\ref{fig:phi_lam100}.
    In the left panel, we show $F'^2=2(B'^2-E'^2)$, indicating electric (red) or magnetic (blue) dominance, and in the right panel, we show 
    $\epsilon_{\mu \nu \lambda \rho}F'^{\mu \nu}F'^{\lambda \rho}=4\mathbf{E}'\cdot\mathbf{B}'$, indicating the degree to which
    the electric and magnetic field are aligned or anti-aligned.
    In both cases, the quantities are normalized by $2\rho_{A'}$, so that they are bounded by $\pm 1$.
\label{fig:vec_lam100}
}
\end{figure}

The vector field configuration post-vortex formation is illustrated in
figure~\ref{fig:vec_lam100}. The oscillating regions of electric dominance (see
figure~\ref{fig:field_comp}) are where the vortex density is
the highest. There is also a strong alignment of the dark electric and magnetic
field in the regions of electric dominance. 

\subsection{Three-dimensional results}
\begin{figure}
\begin{center}
\includegraphics[width=0.6\textwidth,draft=false]{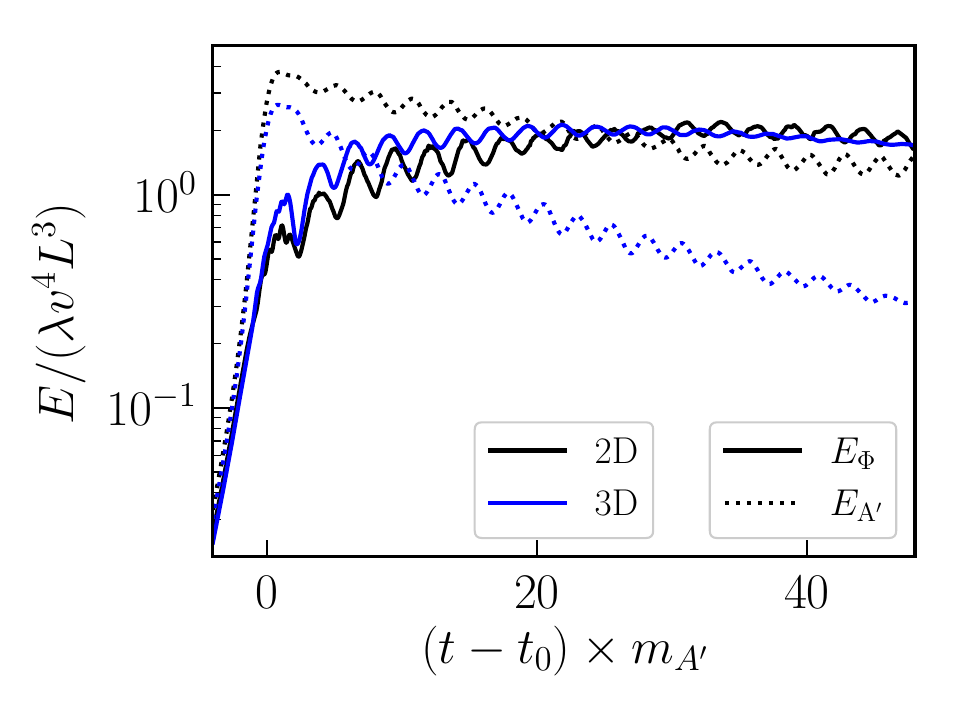}
\end{center}
\caption{
    Energy in the scalar field (including interaction energy with the vector; solid lines) and
    energy in the vector field (not including interaction energy; dashed lines) as a function.
    We compare two cases with $\lambda/g_D^2=25$: one where a translational symmetry is enforced
    in one spatial direction (black lines labelled 2D), and one where on symmetries are enforced
    (blue lines labelled 3D).
\label{fig:e_2d3dcomp}
}
\end{figure}

Though the above results were obtained assuming a translational symmetry in one
spatial dimension (e.g. string vortices are infinitely long), we also perform
fully 3D calculations and confirm that our main findings on vortex formation
are independent of this assumption.  Specifically, we consider a fully 3D case
with $\lambda/g_D^2=25$.  We compare this to the equivalent case assuming a
translational symmetry in figure~\ref{fig:e_2d3dcomp}. There it is evident that
the evolution of the energy in the scalar and vector components is similar in
the two cases, with the only difference being a faster decay in the vector
field energy for the fully 3D case. Though partially a function of finite
numerical resolution, a more efficient transfer of energy from the vector to
scalar field in 3D is expected since (see section~\ref{sec:electric}), with the
symmetry assumption, the dark electric field can only accelerate strings in
the transverse direction, while without the symmetry assumption, the dark
electric field can accelerate strings and the dark electric and magnetic
fields can stretch the length of the strings. 

\begin{figure}
\begin{center}
\includegraphics[width=0.44\columnwidth,draft=false]{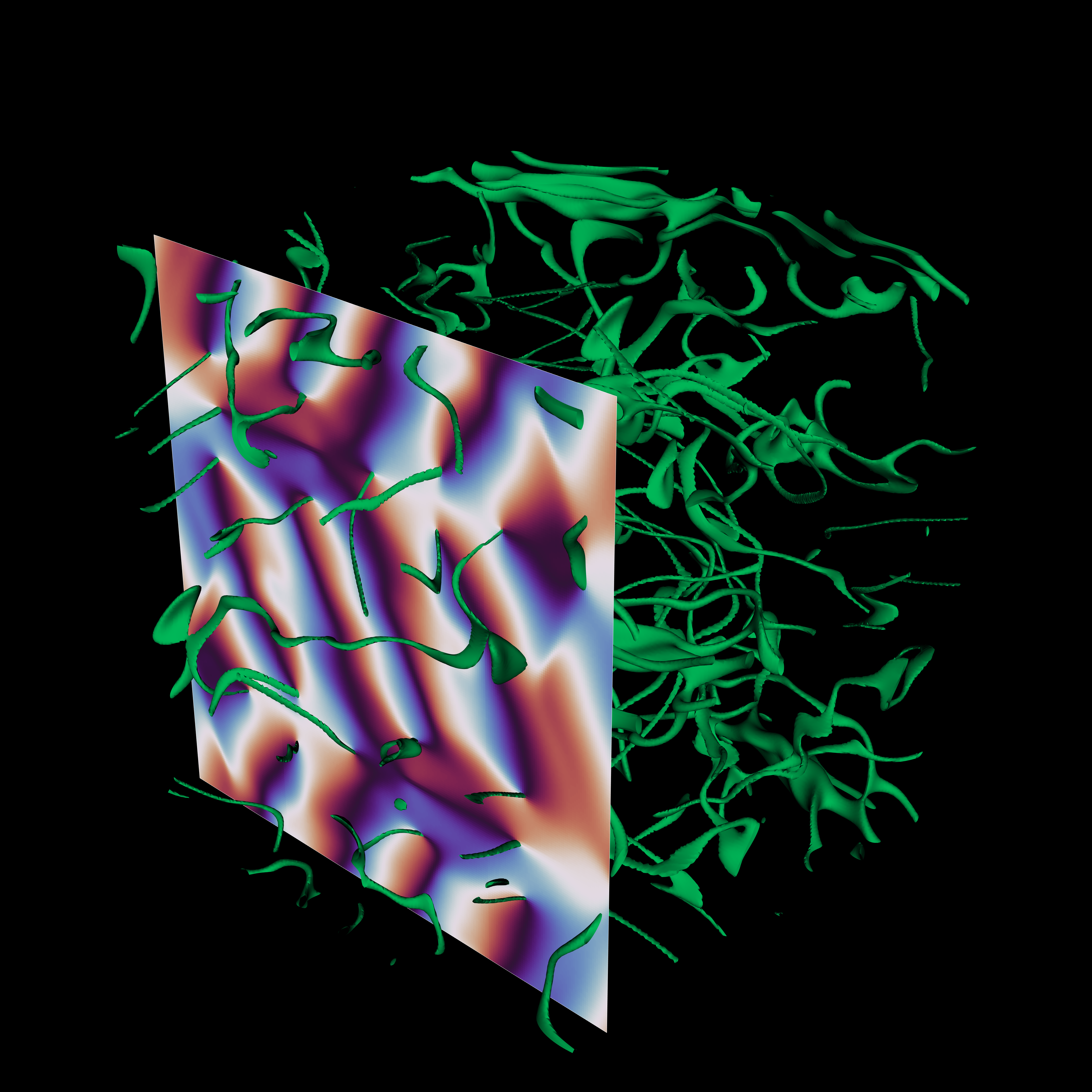}
\includegraphics[width=0.44\columnwidth,draft=false]{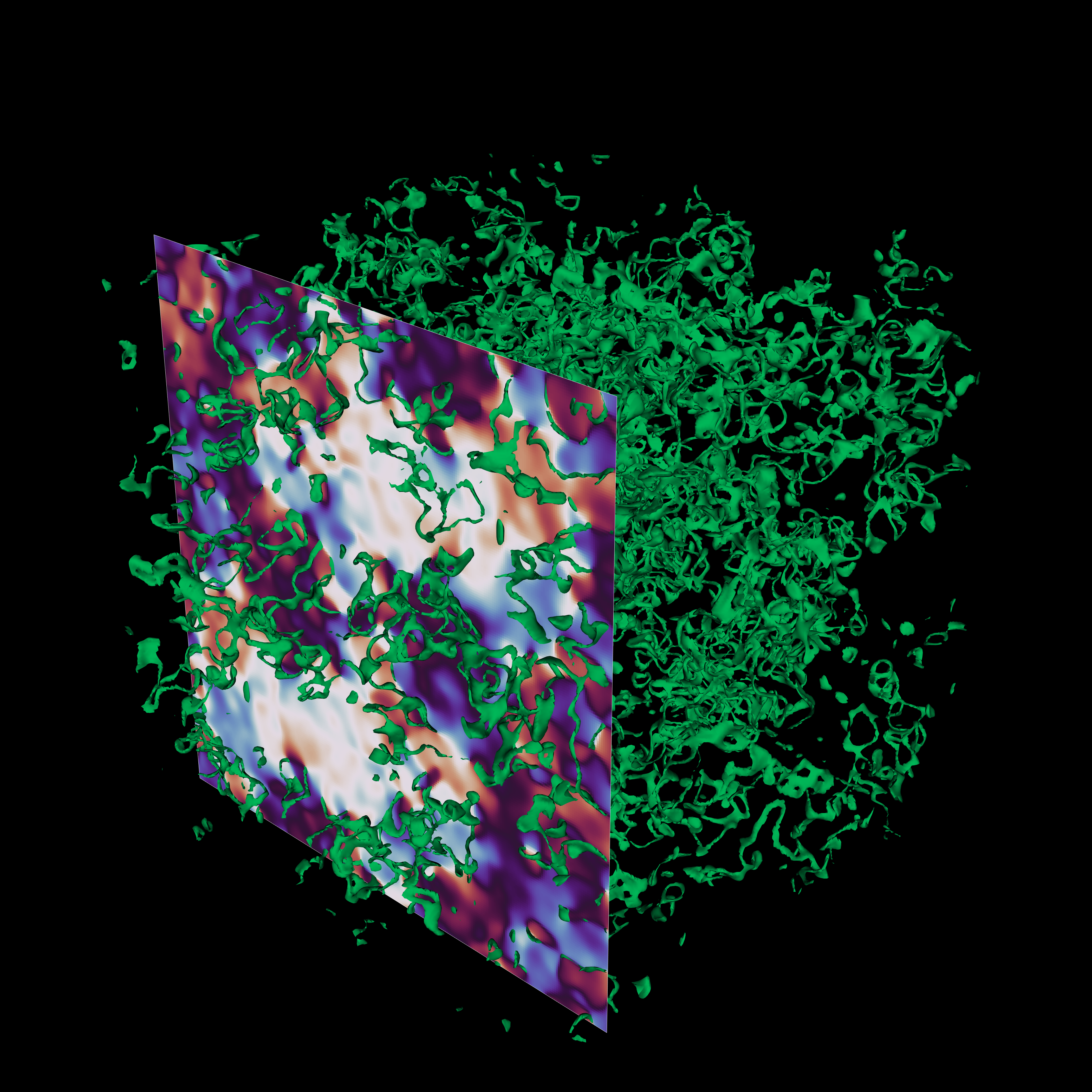}
\raisebox{1.0cm}{\includegraphics[width=0.1\columnwidth,draft=false,trim=200 0 0 0, clip]{argphi_colorbar}}
\end{center}
\caption{
Scalar field from a fully 3D dimensional case with $\lambda/g_D^2=25$.
Contours of $|\Phi|=v/10$ are indicated in green, while the complex phase of $\Phi$ is indicated on a representative two dimensional slice.
The left panel shows a snapshot $\approx 3m_{A'}$ after vortices first form. 
The right panel shows a snapshot at a time $\approx48m_{A'}$ after the first,
illustrating how reconnections have lead to a larger number of smaller scale loops. 
The length of the domain is the wavelength of the dominant mode of the axion instability, 
which is approximately equal to the Compton wavelength of the dark photon ($L=2\pi/k_p$ where
$k_p \approx 1.055 m_{A'}$).
\label{fig:lam25_3d}
}
\end{figure}

Another difference we find, when not assuming a spatial symmetry, is that
intersections between vortex strings efficiently lead to the formation of a
large number of smaller scale closed loops, a sign of a fast melting phase
transition, analogous to the one discussed in section~\ref{sec:melting}, with the mild $\lambda/g_D^2$ values used in the simulation. 
This is illustrated in figure~\ref{fig:lam25_3d}. Also, in contrast to
figure~\ref{fig:phi_lam100}, regular dense bundles of vortices are not evident
at late times. 

We find that vortices form when the magnetic field reaches $B'^2 \sim 2 \lambda
v^4$.  This field strength is comparable to the superheating field strength in
the Landau-Ginzburg theory, suggesting that the superheating phenomenon occurs
regardless of if the initial condition is the Meissner phase, where magnetic
field is absent in the bulk, or dark photon dark matter, where the magnetic
field already exists in the bulk. A possible explanation is that even in the
Meissner phase, within a magnetic penetration depth $1/m_{A'}$ from the edge of
the superconductor, there is magnetic field with comparable strength to the
external magnetic field, and the existence of edges is also not required for
the formation of vortices. 

\section{Evolution post vortex formation}\label{sec:evo}
Once vortices form due to the production  of a longitudinal or magnetic mode,
they will evolve inside the background electromagnetic field. In the case of a
type II superconductor, a two-dimensional Abrikosov lattice forms in the
background of a uniform, constant magnetic field, and the magnetic field gets
confined inside these vortices. In the context of dark matter, this would correspond to moving the energy that
is stored in the background dark photon field into the energy of
the vortices, which subsequently radiates away a sub-component of their energy into
dark photons, and possibly gravitational waves, of much higher energy,
depleting the dark photon dark matter. 

The post vortex formation evolution of dark photon dark matter differs
conceptually from the superconductor case in two ways.  Firstly, as mentioned
in the context of vortex production, there is not really an edge of the
superconductor in the case of dark photon dark matter. Secondly, the strings
start in random directions at the time of formation on large scales, but are
aligned on small scales. In the case of longitudinal mode production during
inflation, these dark photon strings evolve in a manner that is very similar to the case of
gauge strings.  However, in the case of magnetic mode production in the late
universe, there will be a superheating period during the production of magnetic
modes, which ends with the rapid formation of strings with length per Hubble
patch much larger than the density of a scaling solution. This sets off a period
of rapid evolution, during which there is rapid reconnection, and a burst of
dark photon and gravitational wave emission.

At formation, assuming that $m_{\rho} \gg m_{A'} \sim k_c \sim H$, 
the vortices form as roughly parallel lines with distances of $1/k_c$, which evolves
eventually to $a_L \sim \left(m_{\rho} m_{A'}\right)^{-1/2}$ and the energy
density of these vortices in their ground state is
\begin{equation}
    \rho_v \sim v^2/a_L^2 \simeq m_{\rho} m_{A'} v^2,
\end{equation}
whereas the energy density in the dark photon field before vortex formation is
$B^2 \sim v^3 m_{\rho}\gg \rho_v$, as expected from the superheated nature of
the phase transition. On the other hand, the total energy stored in the
electric field and magnetic field before the phase transition, even if all
converted into thermal energy of the radial mode, can barely reheat the system
enough to totally restore the symmetry. Therefore, the evolution of the string
network post vortex formation is determined by the effective theory of the vortices
interacting with a background dark electromagnetic field. 

\subsection{Dark electric field in the presence of vortices}\label{sec:electric}

\begin{figure}
    \centering
    \includegraphics[width=0.5\textwidth]{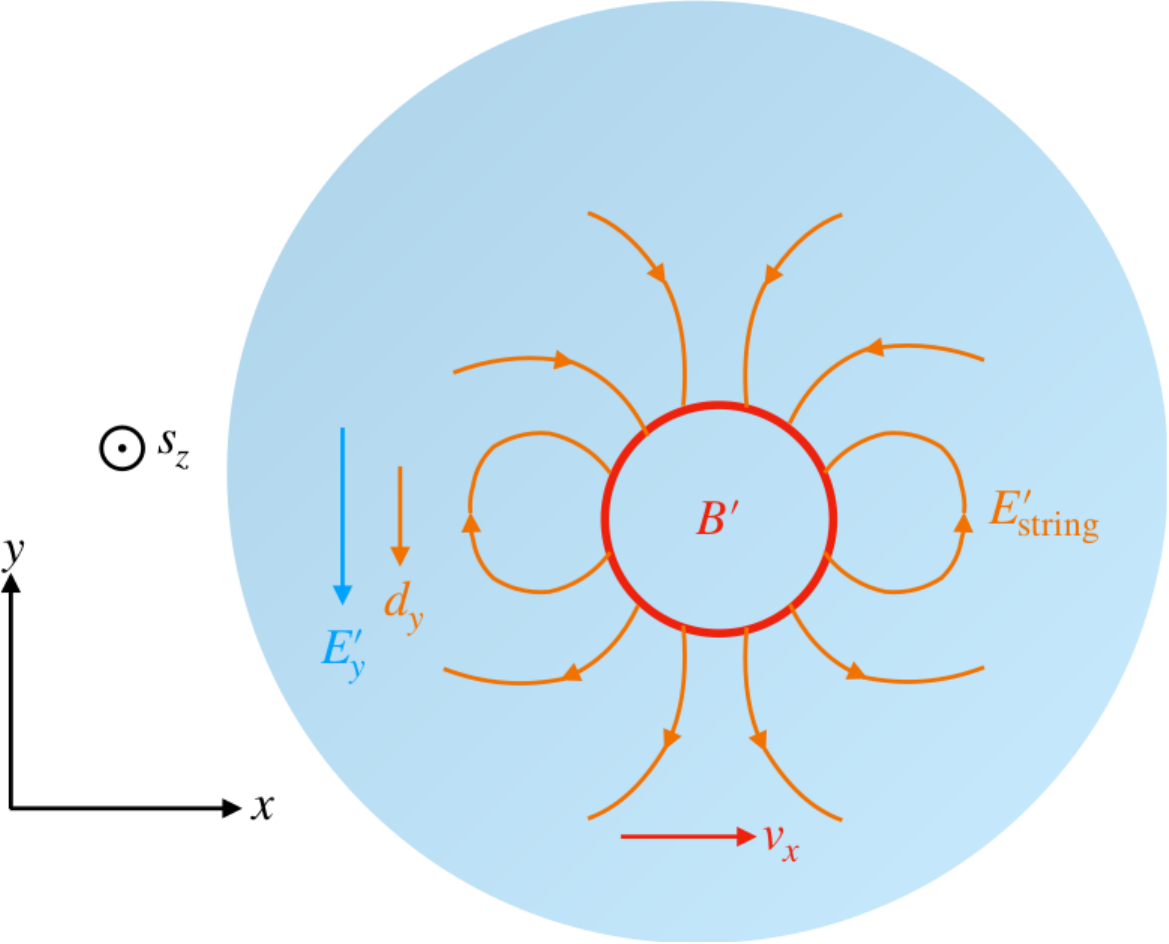}
    \caption{A vortex aligned in the $\vect{z}$-direction accelerates inside a
    background dark electric field $\vect{E}'$. The moving vertex (red) has a
    dipole electric field from a dipole moment of $\vect{d} = - v_x\Phi_D
    \hat{\vect{y}}$ (orange). The velocity, and hence the dipole moment,
    increases to reduce the dipole energy $-\vect{d}\cdot \vect{E}'$.}
    \label{fig:Dipole}
\end{figure}

When vortices form in a background electromagnetic field, the magnetic
field, and hence the magnetic field energy, is concentrated in the vortices,
resulting in a highly excited state of vortices. However, in the case of dark
photon dark matter, there might still be a significant amount of energy 
stored in the oscillating dark electric field. In the following, we will
give a simple description of the interaction of the vortices with the background electric field and
how the electric field is dissipated. To begin, let us consider a vortex line
pointing in the z-direction moving in the x-direction with velocity $\vect{v}=v_x \hat{x}$, as shown in
figure~\ref{fig:Dipole}. The dark electric field outside of the core can be computed
as
\begin{equation}
    \vect{E'} = \frac{d \vect{A'}}{d t} = - \vect{v} \cdot \nabla \vect{A'}
\end{equation}
where the spatial vector $\vect{A}'$ comes from the vortex
\begin{equation}
    \vect{A'} = \frac{1}{2 g_D} \frac{\hat{\vect{\theta}}}{r}
\end{equation}
and as a result
\begin{equation}
    \vect{E'} = - \left(\frac{v_x\Phi_D}{2 \pi r^2}\right)\left(\hat{\vect{r}} \sin \theta - \hat{\vect{\theta}}\cos{\theta}\right)
.
\end{equation}
This is the electric field of a dipole in two dimensions with dipole moment of
$\vect{d} = - v_x\Phi_D \hat{\vect{y}}$, where $\Phi_D = \pi/g_D$ is the dark magnetic flux quantum. As a result, a vortex moving in an
electric field in the y-direction, experiences a force in the same direction,
and energy is transferred from the oscillating electric field to the kinetic
energy of the vortex. During subsequent collisions of the vortex lines, this
kinetic energy can be transferred to potential energy (string length) and the
dark electric field energy is dissipated.
This is evident in the superradiance simulation in~\cite{East:2022ppo} (discussed in section~\ref{sec:vortexpheno}), as well as the 3+1d simulation discussed in the previous section.

The vortex dynamics discussed in this section can also be seen easily
from the perspective of the particle-vortex duality in
$2+1$D~\cite{Peskin:1977kp}, where the electric field in the x-direction is
dual to a current in the y-direction in the XY model. In the dual theory, the
current in the original theory is dual to the gauge field in the dual theory,
and vortices will have velocities of order $E/B$ where $E$ and $B$ are the
electric and magnetic field strength before the superheated phase transition.

\subsection{Residual field}

A moving vortex line aligned in the $z$-direction has a 2D magnetic dipole moment $\vect{\mu} =\Phi_D \hat{\vect{z}} $ and a 2D electric dipole moment $\vect{d} = - v_x\Phi_D \hat{\vect{y}}$. Translating to 3 dimensions, we have
\begin{eqnarray}
    \frac{d \vect{\mu}}{d l} \sim \frac{d \vect{d}}{d l} \sim 1/g_D 
\end{eqnarray}
when a section of string is moving relativistically, where $l$ is the length along the string direction. In a background dark electromagnetic field, a section of string experiences a force from the dark electromagnetic field, as well as from the string tension $\mu \sim \pi v^2 $. The string length, as a result, will increase as long as
\begin{equation}
     \frac{d \vect{\mu}}{d l} \cdot \vect{B'} > \pi v^2, \quad, \frac{d \vect{d}}{d l} \cdot \vect{E'} > \pi v^2,
\end{equation}
when the dark electromagnetic field energy transfers to the energy of the string. Such a process stops when
\begin{equation}
    |\vect{B'}|,\,  |\vect{E'}| \lesssim g_D v^2.
\end{equation}
Note that $g_D v^2$ is exactly $B_{\rm c1'}$, the magnetic field strength when
the Abrikosov vortex solution becomes energetically favorable, as we expect. To
conclude, in a system that does not have a magnetic field direction that is
enforced by an external source and that allows for the dissipation of vortex
energy, the background dark  electromagnetic field will continue to decreases
until it reaches $\mathcal{O}(g_D v^2)$. Such a continues decrease is seen in
both simulations in 3D (figure~\ref{fig:field_comp} and
figure~\ref{fig:field_comp_superrad}). In neither case, however, do we see the field
strength decreases to as low as $B'_{\rm c1} = g_D v^2$ in the duration of the
simulation. This can  be a result of the finite duration of the simulation, as
well as the overlapping magnetic field coming from the magnetic flux carried by
the vortex itself. In the cosmological case, a full simulation of the vortices
with cosmic expansion might distinguish between the two possibilities.  We
should stress that such dynamics is independent of the origin of the
vortex/string and can occur in any growing gauge field in the presence of a
defect. We postpone the study of these scenarios to future work.

\subsection{A super-scaling network}

As described in the previous sections, 
after exceeding the superheating threshold,
most of the coherent dark electromagnetic field gets converted into the string
network. This highly excited string network contains energy density that is
equal to the electromagnetic energy stored in the coherent field, of order
$\lambda v^4$, whereas the energy density of a network approaching the
scaling solution in a cosmological setting has energy $H^2 v^2 \sim g_D^2 v^4 $. This suggests 
that at the end of the phase transition, there is a network of strings with $\xi_i \approx
\lambda/g_D^2\gg 1$, where $\xi_i = \rho_{\rm string}/\mu H^2$ measures the total length of the strings in Hubble units, and would be an $\mathcal{O}(1)$ number for a scaling network. 

This super-scaling network will quickly approach scaling by emitting dark
photons, gravitational waves, and possibly also the radial mode of the Higgs.
The exact composition and spectrum of the radiation depends mainly on the
dynamics of the melting phase transition. In particular, it depends on if the
restoration of the translational symmetry in the direction of the magnetic
field happens at the same rate as the restoration in the directions
perpendicular to the magnetic field.  An $\mathcal{O}(1)$ portion of the vortex
energy can be re-emitted into higher energy dark photons, which would redshift
to be non-relativistic again, and thus constitute a significant portion of the
dark matter. The string network would decay dominantly into high energy
($\omega \gg H\sim m_{A'}$) dark photons if the melting transition is fast. On
the other hand, if such a melting transition is slow, then the network can
decay predominantly into gravitational waves. After the melting phase
transition, the super-scaling network approaches the scaling
solution~\cite{Polchinski:2006ee}. It is not clear how fast this occurs, which
affects the prediction for the gravitational wave and dark photon frequency
distribution around the peak frequency.  We discuss the phenomenological
consequences in section~\ref{sec:gw}.

\section{Dark photon superradiance cloud}
\label{sec:superrad_cloud}
In~\cite{East:2022ppo}, it was demonstrated that string vortices could form as
the results of the superradiant instability of a dark photon around a spinning
black hole. We begin by briefly summarizing those results.  At smaller field
values, the dark photon field acts like a Proca field, and grows exponentially
in time.  Provided the instability does not saturate through gravitational
backreaction beforehand, as $|F'^2|$ reaches $\sim B'^2_{\rm sh}$, string
vortices form. The string formation first occurs with a pair of strings
orientated longitudinally around the black hole. The dark electric field drives
one string into the black hole, and the opposite winding number string outward
(as in section~\ref{sec:electric}).  The latter first grows, with a large
portion of the energy originally in the vector field going in the scalar, in
particular the kinetic (rotational) and potential energy (length) of the
string, and then shrinks, as a combination of the string tension and gravity
cause it to collapse back to the black hole horizon.  Several strings loops 
in the vicinity of the black hole horizon are excited, but also eventually fall back into the
black hole.  Following the transient period where string vortices exist, and a
large fraction of the vector cloud is dissipated due to radiation (as well as
flux into the black hole), superradiant growth begins again. 

\begin{figure}
\begin{center}
\includegraphics[width=0.5\columnwidth,draft=false]{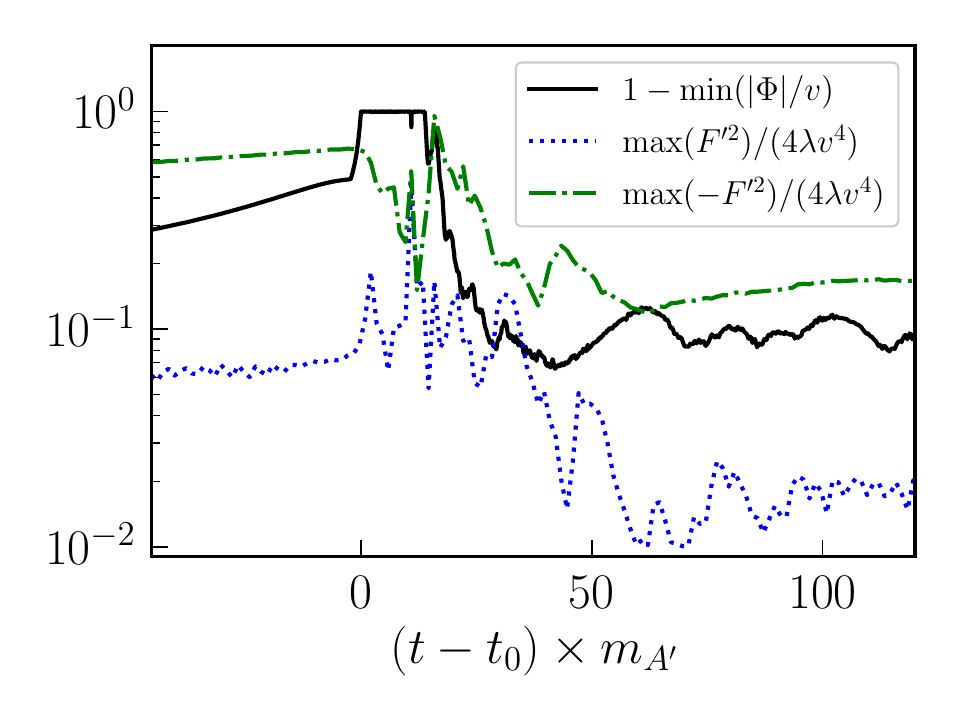}
\end{center}
\caption{
    Similar to figure~\ref{fig:field_comp}, but for a dark photon cloud that
    arises through superradiance around a spinning black hole.  This is the
    case with dimensionless black hole spin $\chi_{\rm BH}=0.99$, $\alpha\equiv G
    M_{\rm BH} m_{A'}=0.4$, and
    $\lambda/g_D^2=25$ from~\cite{East:2022ppo}.  We show the displacement from
    the VEV of the minimum scalar field magnitude (solid black line) around
    the time when vortices form at $t=t_0$.  The vortices persist for
    $\approx14 m_{A'}$ and then are absorbed by the black hole.  We also show
    the maximum values of $F'^2$ (dotted blue line; corresponding to magnetic
    dominance) and $-F'^2$ (dashed-dotted green line; corresponding to electric
    dominance). 
\label{fig:field_comp_superrad}
}
\end{figure}

In figure~\ref{fig:field_comp_superrad}, we show the field strengths from one
of these cases. Comparing to figure~\ref{fig:field_comp}, we can note several
differences in the superradiant instability case versus the axion instability
case.  In contrast to the axion instability case, the Proca cloud is
electrically dominated almost everywhere, and in the regions of magnetic
dominance, the field strength is significantly smaller. Another difference is
that the superradiant instability rate is much smaller than $m_{A'}$ (in
contrast to the axion instability, where they are comparable), and the initial
formation of vortices, marked by the jump of $\min{|\Phi|}$ from $\approx 0.5v$
to $0$ in figure~\ref{fig:field_comp_superrad}) occurs over a short timescale
compared to the instability timescale, which accompanies a similar jump in the
magnetic field strength.  When this jump occurs, the dark magnetic field
strength is below the superheating field strength $B'_{\rm sh}$. This suggests
that the presence of a stronger background electric field can assist the
production of vortices through, we speculate, an instability similar to the
one shown in figure~\ref{fig:Insta}. The relation to the dynamics studied
in~\cite{Bachas:1992bh} and its generalization is unclear.

The characteristic size of the superradiant dark photon cloud is set by $\alpha
\equiv G M_{\rm BH} m_{A'}$ (where $M_{\rm BH}$ is the black hole mass), and is roughly $1/(\alpha m_{A'})$, so the ratio
of the cloud size to the the string radius will be set by $\alpha$ and
$\lambda/g_D^2$.  The study in~\cite{East:2022ppo} was restricted to cases with
$\alpha=0.3$--0.4, and $\lambda/g_D^2=12.5$--50. For these parameters, only a few
string vortices formed (though more were found in the $\alpha=0.3$ versus the
$\alpha=0.4$ case at fixed $\lambda/g_D^2$). However, we can expect that the
dynamics will be much more complicated as one increases the ratio
$\lambda/g_D^2$, and that in particular more strings will be formed as the
scale separation between the radial mode mass and the dark photon mass is
increased, since we do not expect the presence of strings to impede the
formation of new vortices in the superradiance cloud~\cite{Ramos:2005yy}.

Though it is not computationally feasible to numerically simulate the $\lambda/g_D^2 \rightarrow
\infty$ limit, we can estimate the number of vortices that would be formed 
in the superradiance cloud in this regime using the results outlined
in section~\ref{sec:Abrikosov}. 
From there, vortices
are expected to form with a separation on the order of $\sqrt{\pi/g_D B'} \sim
\sqrt{\pi/g_D\lambda^{1/2}} /v$, while the superradiance cloud has a radius of
order $1/\alpha m_{A'} \sim 1/ \alpha g_D v$. Therefore the number of vortices
in the cloud in the limit of large $\lambda/g_D^2$ and small $\alpha$ is roughly
\begin{equation}
N = c_1  \frac{1}{ \pi\alpha} \sqrt{\lambda/g_D^2}/\log[\lambda/g_D^2] \ .
\end{equation}
The coefficient $c_1$ is an $\mathcal{O}(1)$ constant that depends mainly on
the geometry of the system. Note that this scaling might not be very accurate when
$\lambda/g_D^2$ is only moderately large, especially around a black hole with
$G M_{\rm BH} \sim a_L$. 

As an illustrative example, consider a $60\ M_{\odot}$ black hole that is rapidly spinning and
a dark photon with mass $m_{A'}\simeq 10^{-12}$ eV. Then vortex formation will happen 
before superradiance saturates due to black hole spin down if $v \lambda^{1/4} \lesssim 10$ MeV.
This means that the ratio
\begin{equation}
{\lambda/g_D^2} \lesssim \lambda^{1/2} \left( \frac{10\,{\rm MeV}}{10^{-12}\,{\rm eV}}\right)^2 = \lambda^{1/2} 10^{38}
\end{equation}
can be extremely large, in particular if $\lambda$ is order unity.
If, on the other hand, one considers very small $\lambda$ (and even smaller $g_D$), $\lambda/g_D^2$
does not have to be humongous, and the energy scale of $v$ can be much larger.
We note that if one sets the maximum energy density of the dark photon cloud at
gravitational saturation equal to $\lambda v^4$, this gives 
$v \sim \alpha^{5/2}(g_D/\lambda^{1/2})m_{\rm Pl}$, 
independent of the black hole mass.
The different choices for the energy scale $v$ will result in different phenomenology for the vortex
strings, as we discuss below.

\subsection{Evolution post vortex formation in the large ${\lambda/g_D^2}$ limit}\label{sec:SRpost}
For a superradiance cloud that has fully grown, the quantity ${\lambda/g_D^2}$
can be as large as $ 10^{38}$, which corresponds to the formation of
$\mathcal{O}(10^{19})$ vortices around a single solar mass black hole.
Analogous condensed matter systems in the large ${\lambda/g_D^2}$ limit exhibit
a melting phase transition after the superheated phase transition, though it is unclear in our case how fast the phase transition occurs and if all translational symmetries are restored at the same
time. Though such dynamics was not observed for the moderate values of
$\lambda/g_D^2$ considered in~\cite{East:2022ppo}, where only a few strings form in the superradiance cloud,
we can expect similar behavior at more extreme values,
with a large number of strings interacting in the presence
of the time dependent dark electric field, creating loops of irregular sizes.

\begin{figure}
    \centering
    \includegraphics[width=0.8\textwidth]{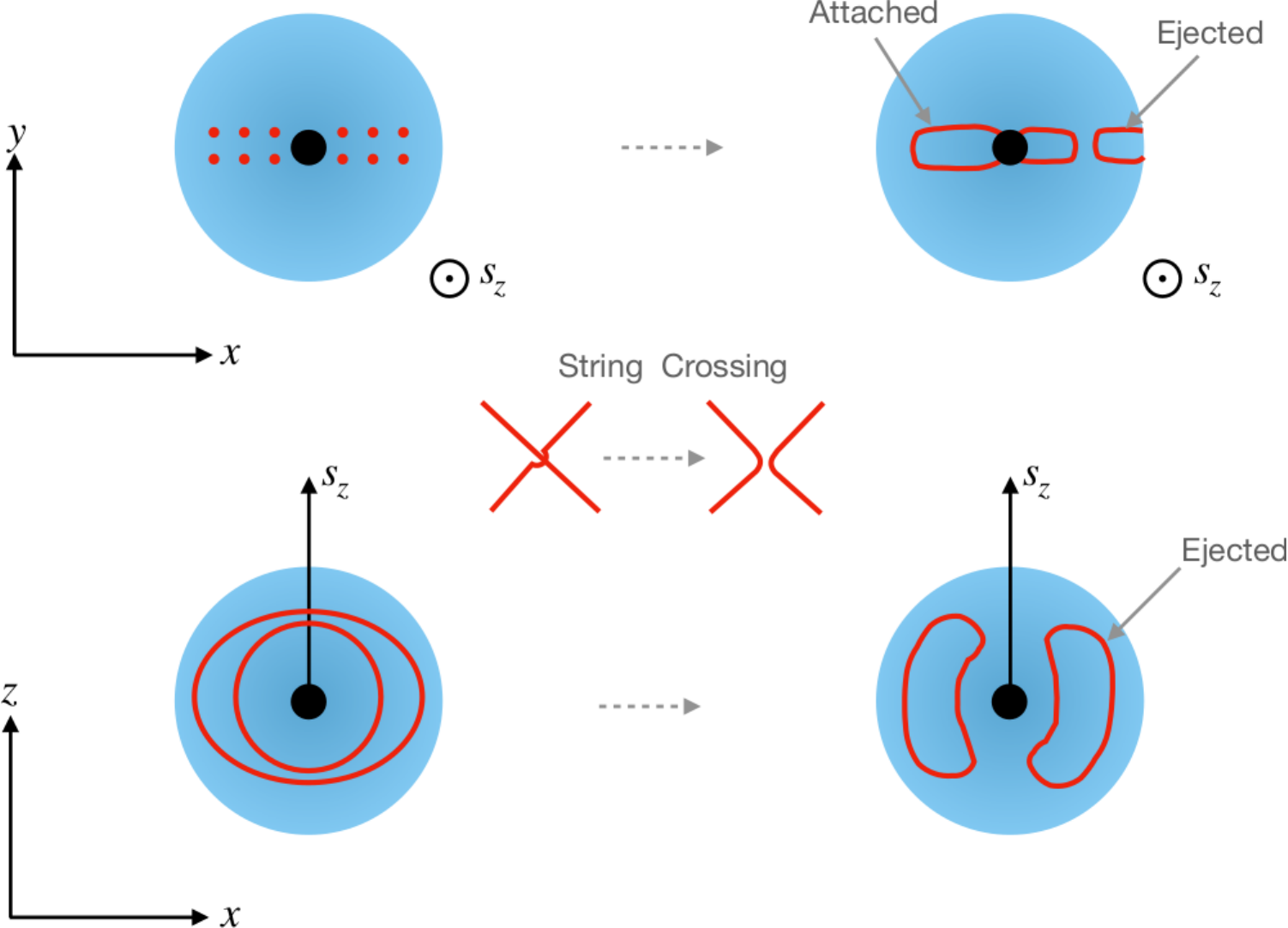}
    \caption{An illustration of how the dynamics of crossings (middle panel) 
affect the motions string vortices in a dark photon cloud that arose through superradiance around 
a spinning black hole. Vortices (red dots and lines) form in the
dark photon superradiance cloud (blue shaded region) around a spinning
black hole (black dot) with spin direction $s_z$. Upper panel ($x-y$ plane):
Initially parallel strings in the vortex lattice exchange partners with each
other, leading to strings that attach to the black hole at different longitudes as well as
strings that are ejected. Lower panel ($x-z$ plane): A different configuration where string loops
spanning a longitudinal line
exchange partners with each other, forming loops that can be ejected from the
superradiance cloud. 
    }
    \label{fig:Cross}
\end{figure}

In a type II superconductor, the melting phase transition eventually completes,
and full translational symmetry is restored. This is also the expectation for dark
photon dark matter. In the dark photon superradiance cloud, however, the
situation is much more complicated. Firstly, the superradiance cloud, and hence
the resulting network of vortices, is localized around the black hole by
gravity, instead of being nearly uniformly distributed over the whole Higgsed
phase. Secondly, most of the energy is initially stored in the dark electric
field, which can eject loops that are created before total
thermalization/melting occurs (see section~\ref{sec:electric}).  In
figure~\ref{fig:Cross}, we sketch how string intersections could create loops
that do not encompass the black hole, with a fraction of these being
accelerated to relativistic velocities by the dark electric field, and becoming
gravitationally unbound, while others fall back into the black hole.

\subsection{Superradiance cloud as a string factory}

The initial superheated phase transition produces 
vortex strings with length on the order of the size of the superradiance cloud
$1/(\alpha m_{A'})$ as demonstrated in~\cite{East:2022ppo}. These vortex loops will expand due both to the
background dark electric field, as well as the repulsive interactions between
the vortices when $\sqrt{\lambda/g_D^2}\gg 1$. These interactions likely make
the vortex lines highly irregular before they can escape the superradiance
cloud. However, due to geometric string-string interactions, string loops
oriented longitudinally around the black hole are also unlikely to 
have lengths that are much longer than $1/(\alpha m_{A'})$
when they escape the region.

Assuming an order one fraction of the total string length escapes the black hole
during a stringy bosenova event and an $\mathcal{O}(1)$ quartic coupling, the total number of string loops with
$L_{\rm loop} \sim 1/(\alpha m_{A'})$ ejected can be 
up to
\begin{equation}
    N_{\rm max} \approx \frac{1}{\pi v^2 L_{\rm loop}}\int |\vect{E}'|^2 d V \simeq \frac{1}{\alpha^2}\frac{\lambda}{g_D^2} \approx 10^{39}.
\end{equation}
if we accounted for the assisted pair production in the larger dark electric field seen in the simulation. If we assume the dark magnetic field crosses the superheating threshold, this $\alpha$-scaling would be $1/\alpha^4$.
These irregular loops can radiate both longitudinal mode dark photons, as well as
gravitational waves, losing energy and eventually circularizing. The
circularized loops can only radiate into gravitational waves with power $P \sim
\Gamma G \mu^2$, and hence have a lifetime
\begin{equation}
    t \simeq \frac{1}{\Gamma G v^2 \alpha m_{A'}}\gg t_{\rm age}
\end{equation}
where $\Gamma$ is an $\mathcal{O}(50)$ constant, $\mu$ is the string tension,
and $t_{\rm age}$ is the age of the universe~\cite{Hindmarsh:1994re}. Here, the main uncertainty comes
from a lack of understanding of the melting transition for a superradiant
cloud, and, as a result, uncertainty regarding the spectrum of the dark photon
strings emitted. 

The strings themselves are capable of
superradiance~\cite{Frolov:1995vp,Kinoshita:2016lqd,Igata:2018kry}, which opens
the possibility that after the the initial burst of string production when the
dark photon cloud grows sufficiently large, the strings can continue to grow in
number/length by extracting energy and angular momentum from the spinning black
hole.  Xing et al.~\cite{Xing:2020ecz} describes a scenario where a string
whose ends are attached to the black hole horizon at different longitudes could
exhibit a type of superradiant instability, where tension modes would be
amplified by successive reflections by the horizon. (Though the treatment there
was non-relativistic, and the role of reconnections and the saturation of such
an instability was left open.) Following a stringy bosenova event, a significant
portion of the irregular strings that fall towards the black hole may intersect
the horizon in configurations that enables them to continue to extract energy
and angular momentum from the black hole. (This is in contrast to what was
found in simulations with a small number of strings~\cite{East:2022ppo}, where
the strings that intersected the black hole had both ends attached at approximately the same
longitude and were not long-lived.) The rate of energy extraction per string is set by the string
tension $\mathcal{O}(v^2)$~\cite{Kinoshita:2016lqd,Igata:2018kry,Xing:2020ecz}.
Such an energy extraction rate is much smaller than the superradiance rate of a
full superradiance cloud. However, given the huge number of strings that are
produced, and subsequently attach to the black hole horizon, it is possible
that such an energy and angular momentum extraction rate is enhanced by the
$\mathcal{O}(\sqrt{\lambda/g_D^2})$ number of strings attached to the black
hole, beyond which point the string cores start to overlap with each other. 

These strings that intersect the black hole horizon might continuously extract
energy from the black hole through string superradiance. However, a visible
signal from string superradiance would require a huge numbers of strings
extracting energy from the black hole at the same time. On the other hand, one
or more strings could impede the superradiant growth of a dark photon, assuming
that the dark photon cloud loses energy by accelerating the strings
(section~\ref{sec:evo}) at a greater rate than it extracts energy from the
spinning black hole.  The phenomenological consequences of these vortices will
be discussed in section~\ref{sec:vortexpheno}. 

\section{Phenomenology}\label{sec:pheno}

The main phenomenological consequences of the dynamics studied in this paper
are the early depletion of dark photon dark matter and dark photon
superradiance clouds through vortex formation. The energy that is transferred to
string networks, however, opens up new phenomenological possibilities. In the
case of vortex formation in the early universe, the dense network of string
would have already decayed away, and we can look for the indirect evidence of
their existence through their gravitational wave signals. In the case of vortex
formation from black hole superradiance, there are direct signals coming from
vortex lines ejected by the black hole. 

\subsection{Gravitational waves from dark photon dark matter}\label{sec:gw}

\begin{figure}
    \centering
    \includegraphics[width=0.6\textwidth]{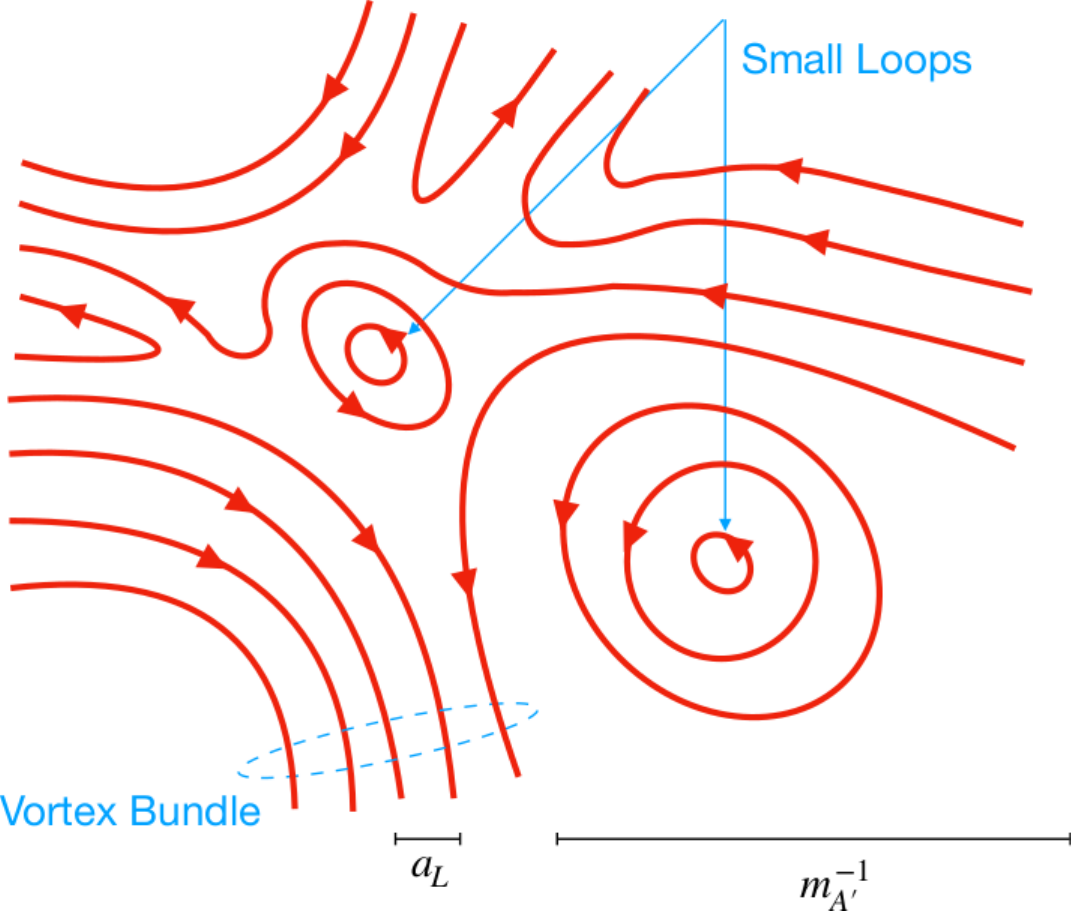}
    \caption{A schematic view of the vortex network generated after a
    superheated phase transition, but before the melting phase transition
    is complete. The long vortex lines have length $L \gtrsim m_{A'}^{-1}$
    and relatively uniform direction on scales of order $m_{A'}^{-1}$, forming a
    vortex bundle. The vortex line separation is of order $a_L$, the lattice
    spacing. The long vortex bundle emits gravitational waves with frequency
    $\omega \sim 1/L$, while the small string loops disappear quickly into high
    energy dark photon longitudinal modes with $\omega \sim 1/a_L$. }
    \label{fig:VortexProfile}
\end{figure}

In the case of inflationary dark photon production, the vortices formed during
inflation will quickly evolve into a scaling solution of cosmic strings after
reheating if the Hubble scale is larger than the scale $v$. If $v\gtrsim H_{\rm
I}$, however, the dynamics can differ from the formation of Nambu-Goto
strings~\cite{NambuGoto}. Unlike Nambu-Goto strings, which can only shrink in
total length after they enter the horizon, the length of dark photon strings
can continue to grow as strings enter nearby Hubble patches which contain a
dark electromagnetic field. This allows the string network to evolve into
scaling even if the original exponentially suppressed production during
inflation cannot produce enough strings to enter a scaling regime. We leave a
more in-depth study to future work.

In the case of late time production mechanisms for the dark photon, vortices are
produced in a superheated phase transition as the dark electromagnetic field
grows. These strings form a super-scaling network, with an effective $\xi_i
\gtrsim \sqrt{\lambda/g_D^2}$. After formation, the string length can continue
to grow inside the background dark electric field to a total length of string
that corresponds to at most $\xi_i \lesssim \lambda/g_D^2$. 

After an initial transient phase, the background dark electromagnetic field is
dissipated, and the strings will continue to evolve via string-string
interactions towards a scaling network. The system differs from the standard
picture in two ways. Firstly, there are a large number of defects in a single
Hubble patch, similar to the string network after a second order phase
transition~\cite{KIBBLE1980183,zurek1985cosmological,Kibble:1997yn}. Secondly,
the network, before the melting phase transition is complete, breaks
translational symmetry and is maximally anisotropic on Hubble scales, and
this anisotropy can be long-lived since the densely packed strings 
have a significant repulsion between them. 

\begin{figure}
    \centering
    \includegraphics[width=0.6\textwidth]{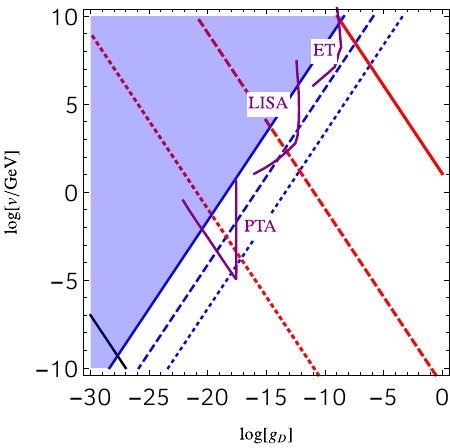}
    \caption{The observational consequence of a network of string bundles that
    radiates predominantly gravitational waves. The blue shaded region is
    excluded by constraints on the gravitational wave energy density from the measurement of the effective number of neutrino species  $N_{\rm eff}$~\cite{Planck:2018vyg},
    corresponding to dark photon dark
    matter overclosing the universe before the dynamics we discuss here
    take place. The blue dashed and dotted lines are $N_{\rm eff}$ of
    $10^{-5}$ and $10^{-10}$. The red solid, dashed, and dotted lines show the
    peak frequency of the gravitational wave of kilohertz, millihertz, and
    ${\rm yr^{-1}}$, respectively. We also show corresponding near term projections
    for some future gravitational wave detectors (ET, LISA, and a pulsar timing array with SKA, respectively) as solid purple
    curves~\cite{Moore:2014lga}. Below the black solid line, the dynamics we
    discuss here occurs after matter-radiation equality.}
    \label{fig:GW}
\end{figure}

\begin{figure}
    \centering
    \includegraphics[width=0.6\textwidth]{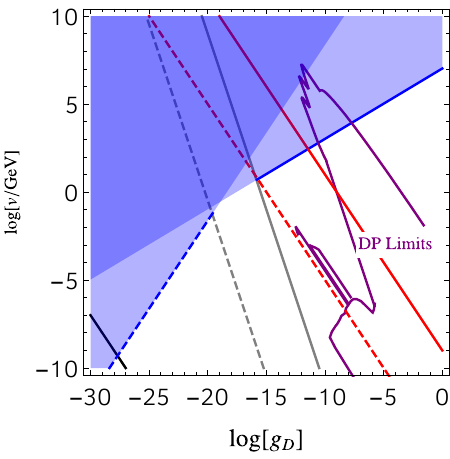}
    \caption{The observational consequences of a network of string bundles that 
    radiates predominantly longitudinal mode dark photons. The blue shaded
    region is excluded by overclosing the universe on the right, and $N_{\rm
    eff}$ constraints on the left~\cite{Planck:2018vyg}. The blue solid line is when dark
    photons make up all of dark matter, while the blue dashed line
    corresponds to when relativistic dark photons have the same energy density as
    photons today. The red solid and dashed lines correspond to when the mass of the dark
    photon is ${\rm eV}$ and ${\rm GHz}$ ($\sim10^{-6}$ eV), respectively. The gray solid and
    dashed lines show when the dark photon radiation becomes non-relativistic at, respectively,
    cosmic temperatures of $T_{\gamma} = {\rm keV}$ and today. In the region between the gray solid
    and dashed lines, dark photon radiation becomes non-relativistic during
    the time of CMB and structure formation, and the constraint depends on the
    details of the spectral shape of dark photons emitted by the string
    network. Below the black solid line, the dynamics we discuss here
    occur after matter-radiation equality. The parameter space to the bottom right of the purple solid line is excluded by the
    existing solar, CMB, and ADMX constraints~\cite{AxionLimits} if the
    kinetic mixing is generated at one loop as in~\cite{Holdom:1985ag}, i.~e.~ $\epsilon = e g_D/16\pi^2$.}
    \label{fig:DarkPhoton}
\end{figure}

The phenomenological consequences mainly depend on whether the emission from the
string network is determined by the behavior of individual string loops, or by
the collective behavior of bundles of strings (see figure~\ref{fig:VortexProfile} for a schematic view of the string network). 
The bundle of long strings will emit significant gravitational radiation at a frequency
corresponding to the inverse of the length of the strings (comparable to Hubble scale at that time). On the other hand, if the
melting transition takes place efficiently after the formation of the strings,
then most of the energy will go into longitudinal dark photons, which have a
frequency of order $1/a_L$ (inverse lattice spacing) as long as $1/a_L$ is
larger than the dark photon mass. It is unclear how this network approaches a
scaling solution, and as a result, our projections for the emission, shown in figures~\ref{fig:GW}
and~\ref{fig:DarkPhoton}, should only be regarded as crude estimates.  

In both figure~\ref{fig:GW} and figure~\ref{fig:DarkPhoton}, we assume that there is
no initial seed of strings produced in the early universe, and that the length of
cosmic string starts decreasing rapidly after the phase transition. The former
assumption is equivalent to $H_I \ll v$, since the growth of a dark photon dark matter field
will be halted by the presence of seed vortices. The latter is a good
approximation when the string network forms quickly, and dissipates its energy
mainly into longitudinal dark photons at $\omega \sim 1/a_L$
(see~\cite{Long:2019lwl} for more details). In this case, most string loops,
after the melting phase transition, have length in the range
\begin{equation}
 \frac{1}{m_{\rho}} \ll L_{\rm loop}  \ll \frac{1}{m_{A'}}
\end{equation}
and the behavior of the string network is the same as an axion string network
until $\xi_i$ drops to be $\mathcal{O}(1)$. On the other hand, if the string
network dissipates a large portion of its energy before the melting phase
transition is complete, then it is possible that the energy released into
gravitational waves is low. The main uncertainty in estimating the gravitational
wave power comes from the effect of interference. If the bundle of 
strings emit gravitational wave independently, then the emission power
scales linearly with $\xi_i$, and the lifetime of strings with length
$L > 1/m_{A'}$, 
much longer than the Hubble scale when the string network forms,
is $\mathcal{O}(L/G v^2)$ due to gravitational wave
emission. 
In this case, the string network will predominantly emit dark photons after the
melting transition is complete. However, if a bundle of strings
oscillates and emits gravitational waves coherently before the melting phase
transition is complete, then the emission power scales quadratically with
$\xi_i$, and the lifetime of long strings is $\mathcal{O}(L/G \xi_i v^2 \sim
L)$ due to gravitational wave emission, and $\mathcal{O}(1)$ of the energy can
be released before the melting phase transition is complete. Simulations of a
super-scaling network of gauged strings with anisotropic initial conditions
would help make a more precise prediction of the phenomenological
consequences.

In both figures, we assume the energy density that is transferred into
gravitational wave or dark photon dark matter radiation is
$\mathcal{O}(\lambda v^4)$ at the phase transition, which happens when vortices
form quickly, and shut off the exponentially growing gauge field. It is
possible that a small amount of energy is continuously pumped into the dark
gauge field post string formation. In this case, there might be a significant
blue tail to the gravitational wave spectrum since the correlation length of
the magnetic field is $\mathcal{O}(1/m_{A'})$, which remains relatively
constant until the melting transition is complete.

\subsection{Stringy bosenova}\label{sec:vortexpheno}
\subsubsection{Dark magnetic flux tube} 
\begin{figure}
    \centering
    \includegraphics[width=0.6\textwidth]{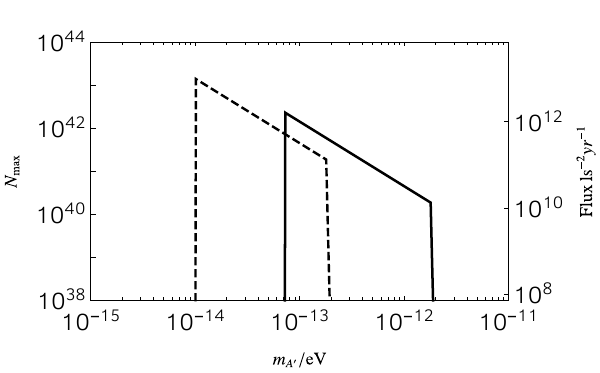}
    \caption{Maximum number of vortex lines (left axis) and equivalent flux (in units of inverse
    light seconds squared per year, right axis) 
    emitted in a stringy bosenova as a
    function of the dark photon mass. The solid and dashed lines correspond to
    black holes with masses of $10 $ and $100 M_{\odot}$, respectively, with 
    dimensionless spin $\chi_{\rm BH}=0.5$ when the stringy bosenova event happens. We take the quartic $\lambda =1$,
    and assume that the dark photon superradiance grows until the black hole spins
    down by $\Delta \chi_{\rm BH}=0.1$ for the azimuthal number $m=1$ mode, and that the black hole was formed
    $10^{6}\,{\rm years}$ ago.
    }
    \label{fig:VortexNum}
\end{figure}

Unlike vortex formation in the early universe, where an indirect relic
gravitational wave signal can be produced, in the case of a dark photon superradiance cloud,
one can hope to directly observe vortex lines that are ejected from the black hole.
Taking dark photon masses corresponding to superradiance of black holes with masses in the range $10$ to 100 $M_{\odot}$
(see figure~\ref{fig:VortexNum}),
assuming $\lambda=1$, and taking $v$ to be the maximum value where vortex formation happens before
black hole spin down halts superradiance (i.e. setting $\lambda v^4$ equal to the maximum energy density of the cloud), 
we arrive at the following parameteric scales.
Each vortex loop has a core size $1/(\lambda^{1/2} v)$ of order ${\rm pm}\sim
{\rm fm}$, a thickness ($1/{ m_{A'}}$) (region
containing magnetic field) of order $10 \sim 10^{4}\, {\rm km}$, and a length
($\sim 1/{\alpha m_{A'}}$) of order $10^2 \sim 10^{6} \, {\rm km}$. 
We note that for lower values of $\alpha \lesssim 0.1$, it is possible that 
energy emission from nonlinear interactions halts the superradiant growth of
the cloud before the superheating field strength is every reached~\cite{Fukuda:2019ewf,East:2022ppo}.
However, numerical simulations are required to locate the threshold value of $\alpha$ below
which this occurs. 

A fraction of these loops, after escaping from the superradiance cloud, diffuse into the galaxy. 
The flux of dark photon strings from a black hole in the galaxy is
\begin{equation} F_{\rm max} = \frac{N_{\rm max}}{4\pi d^2 \Delta t} \approx
    \frac{10^{10}}{A_{\rm string}} {\rm yr}^{-1}\left(\frac{10\,{\rm
    kpc}}{d}\right)^2 \left(\frac{10^5\,{\rm yr}}{\Delta t}\right),
\end{equation} where $A_{\rm string} = \alpha^{-1} m_{A'}^{-2}$ is the
effective area of the vortex line, $\Delta t$ is the range of arrival times/duration
of semi-relativistic string loops from a distance $d \sim 10 \,{\rm kpc}$, and
$N_{\rm max}$, the maximal number of strings diffusing out of the black hole
superradiance cloud in each bosenova event, is shown in
figure~\ref{fig:VortexNum}. This is an enormous flux. In fact, the magnetic
field from different strings will still overlap after the strings
have diffused from the kilometer-scales of the black hole superradiance cloud to the kiloparsec-scales
of the galaxy. After the initial burst, string superradiance~\cite{Xing:2020ecz} may
continue to produce string loops out of the black hole. This steady flux can be
as large as
\begin{equation}
    F_{\rm steady} \sim \frac{N_{\rm max}}{4\pi d^2 t_{\rm age}} \approx \frac{10^{5}}{A_{\rm string}} {\rm yr}^{-1}\left(\frac{10\,{\rm kpc}}{d}\right)^2,
\end{equation}
if the source black hole is within our galaxy and a large number of strings are attached to the black hole.

Each vortex carries one flux quantum $\Phi_D = \pi/g_D$. If the
dark photon is kinetically mixed with our photon, such a dark magnetic flux
will be a visible magnetic flux of $\Phi_{\rm visible} = \pi
\epsilon/g_D$ where $\epsilon$ is the dimensionless kinetic mixing parameter. The effective magnetic field around a string is
\begin{equation}
    B = \frac{\Phi_{\rm visible} }{\pi/m_{A'}^2} = \epsilon m_{A'} v \simeq 10^{-15} \,{\rm T}  \left(\frac{\epsilon}{10^{-7}}\right)\left(\frac{m_{A'}}{10^{-12}\,{\rm eV}}\right)^{13/4}\left(\frac{M_{\rm BH}}{10\ M_{\odot}}\right)^{7/4}.
\end{equation}
A string that is moving at close to the speed of light would intersect the
detector for $\mathcal{O}(10^{-2})$ seconds. Current magnetometers can be used to
look for such a transient signal~\cite{PhysRevLett.89.130801}. 

If the kinetic mixing parameter is generated at one loop order, then
\begin{equation}
    \frac{ \Phi_{\rm visible}}{\Phi_0} = \frac{ e \epsilon}{g_D} = \frac{ e^2 g_D}{16\pi^2 g_D} = \frac{\alpha_{\rm EM}}{4\pi},
\end{equation}
where $\alpha_{\rm EM} = e^2/(4\pi) \approx 1/137$ is the fine structure
constant. Such a quantised magnetic flux, spread over an area of order ${\rm
km^2}$, has magnetic field strength of order $10^{-26} \,{\rm T}$, and is
beyond the sensitivity of current experiments. One way to enhance the signal is
to have a string that carries a large number of magnetic flux quanta. This is,
however, very unlikely, since the force between fluxes are repulsive in the
limit of large $\lambda/g_D^2$. 

Finally, in passing, we note that
the large number of vortices, if formed and ejected from the black hole with
relativistic velocities, will inevitably travel from galaxy to galaxy, and be
captured by black holes in different galaxies~\cite{Xing:2020ecz}.  This has an
amusing effect of ``infecting" the black holes the strings encounter.  A vortex
string that is attached in a superradiating mode could then hinder vector
superradiance as long as it is attached. 
In certain cases, this means that
vector superradiance might not be occurring
due to the existence of these persisting strings, as light and as insignificant as they
may be compared to the black hole they are attached to~\footnote{We thank Neal Dalal for sharing this
amusing point.}. 
These strings might track their origin back to an
earlier stringy bosenova event, or some early universe vortex
production through either Kibble mechanism~\cite{KIBBLE1980183}, or the
mechanisms we discussed in this paper. We leave a detailed study of this
scenario to future work.

\subsubsection{Gravitational wave sirens}
If we assume the condition for the assisted superheated phase transition $|\vect{E}'|\simeq
B'_{\rm sh}$ (as seen in seen in the superradiance simulations) is satisfied just below when the dark photon
cloud reaches the maximum density allowed by black hole spin down, then we have  
\begin{equation}
   |\vect{E}'|^2 \simeq \alpha^5 m_{A'}^2/G = \lambda v^4 \rightarrow \frac{\lambda}{g_D^2} \simeq \frac{\alpha^5}{ G v^2}.
\end{equation}
Thus strings with tension as large as $ G \mu=\pi \alpha^5 g_D^2/\lambda$ can be produced
and ejected from the black hole in a stringy bosenova event (up to logarithmic
corrections)\footnote{This $\alpha$-scaling would be $\alpha^7$ if we required
that $|\vect{B}'|\simeq B'_{\rm sh}$.}. We consider here the case where the quartic coupling $\lambda$
is very small, and $\lambda/g_D^2$ is only moderately large.
For $\alpha\approx 0.3$ and $\lambda/g_D^2\sim 10^3$,
\begin{equation}
    G \mu \simeq 10^{-5} \left(\frac{\alpha}{0.3}\right)^{5} \left(\frac{10^3}{\lambda/g_D^2}\right),
\end{equation}
suggesting that the stringy bosenova event can lead to the production of strings
that are impossible even in the very early universe. In fact, CMB measurements
suggest that strings produced in the early universe must have tension $G \mu
\lesssim 10^{-7}$~\cite{Charnock:2016nzm}. However, since a stringy bosenova
event only produces very short string loops at late times, none of the strong limits
for the early universe apply.

After production, we expect a fraction of the string loops to be ejected from the black
hole, emitting radiation at first predominately in gravitational waves, and subsequently in dark photons. The
lifetime of such a string loop is
\begin{equation}
    t_{\rm life} \sim \frac{1}{\Gamma G \mu \alpha m_{A'}} \sim \frac{\lambda/g_D^2}{\Gamma \alpha^6 m_{A'}} = 6 \,{\rm kyr} \left(\frac{\lambda/g_D^2}{10^{13}}\right) \left(\frac{0.3}{\alpha}\right)^{6} \left(\frac{10^{-12} \,{\rm eV}}{m_{A'}}\right).
\end{equation}
Two qualitatively different regimes exists: one where the strings have
a lifetime much shorter than the travel time between the source and the earth,
and one where the lifetime is much longer. 
For the former, all
$N_s = \lambda/(g_D^2 \alpha^2)$ strings decay close to the source
black hole, which produces a burst of gravitational wave with duration $t_{\rm
life}$, characteristic frequency $\alpha m_{A'}$, terminal frequency $m_{A'}$
(when the dark photon longitudinal mode emission dominates the energy loss) and
roughly a strain 
\begin{equation}
    h \sim \frac{1}{m_{A'}^{3/2} t_{\rm life}^{1/2} d}  = 4\times 10^{-18} \left(\frac{\alpha}{0.3}\right)^3 \left(\frac{1000}{\lambda/g_D^2}\right)^{1/2} \left(\frac{10^{-12} \,{\rm eV}}{m_{A'}}\right)\left(\frac{10 \, {\rm kpc}}{d}\right) 
\end{equation}
where $d$ is the distance from us to the source. On the other hand, if the
strings have a lifetime much longer than the travel time between the source of the
string and the earth, then the string loops behave as gravitational sirens,
flying around in the galaxy and possibly the universe while emitting gravitational waves. For the string that gets
closest to the earth en route within an observing time $t_{\rm
obs}\ll t_{\rm life}$, the typical distance between it and the earth is $D_s \sim
\sqrt{ d^3/N_s t_{\rm obs}}$. This string generates a strain of
\begin{equation}
    h \sim \frac{\Gamma^{1/2} G \mu }{\alpha m_{A'} D_s} = 3 \times 10^{-26} \left(\frac{\alpha}{0.3}\right)^3 \left(\frac{10^{12}}{\lambda/g_D^2}\right)^{1/2} \left(\frac{10^{-12} \,{\rm eV}}{m_{A'}}\right)\left(\frac{10 \, {\rm kpc}}{d}\right)^{3/2} \left(\frac{t_{\rm obs}}{10\,{\rm year}}\right)^{1/2}.
\end{equation}
Such a signal has a much longer duration, and signals from strings from
different black holes can overlap in time. In the limit $ t_{\rm life} \gg
t_{\rm age}$, the universe is filled with small string loops with density $N_s
n_{\rm BH}$, emitting gravitational waves at the same time. Here $n_{\rm BH}$
is the number density of black holes that have underwent a stringy bosenova event.
Gravitational lensing can also be used to look for these light massive
objects~\cite{Dai:2019lud,VanTilburg:2018ykj}. In addition to the gravitational
wave emission, the string loops also produces a large amount of dark radiation
in the form of longitudinal dark photons. We leave a more detailed study to
future work. Lastly, given the large tension, these strings could also extract
energy from black holes much more efficiently through string
superradiance~\cite{Xing:2020ecz}, which can potentially lead to strings that
are longer than the ones we considered in this section getting ejected from the
black hole.

\section{Remarks}\label{sec:remarks}
In this paper, we point out how the formation of string vortices and the
ensuing nonlinear dynamics can significantly affect the validity of various
dark photon dark matter production mechanisms. These dynamics, however,
introduce new phenomenological opportunities. The direct and indirect signals
from vortex formation and evolution in both dark photon dark matter and dark photon
superradiance clouds can be looked for with gravitational wave detectors, dark
photon detectors, as well as magnetometers.

It is evident that the relative strength of the main observational signals of
our study depends strongly on understanding the melting phase transition, that
is the transition to a large number of vortex strings that are uncorrelated on
large scales, which has been an active research area of condensed matter
physics
~\cite{zeldov1995thermodynamic,james2021emergence,liu1991kinetics,PhysRevB.8.3423}.
A better understanding of how this phase transition takes place following a
superheated phase transition in the large $\lambda/g_D^2$ limit will be
important for narrowing down the uncertainties. 

In a large portion of the parameter space for dark photon dark matter where $v
\lesssim H_I$ or $v \lesssim T_{\rm RH}$ (the reheating temperature), we would
be left with a large number of strings from prior phase transitions, as well as
the dynamics described in this paper. The existence of these strings can impede
the growth of coherent field in the late universe. In the background of these
defects, coherent fields growing past the critical field $B'_{\rm c1} = g_D
v^2$ (and electric field of a similar order) can already be damped by the
strings. Understanding the details of this effect, which we leave to future work, can be important in the context of vector black hole
superradiance, dark matter dynamics in galaxy mergers, and the constraints on
the photon mass~\cite{Adelberger:2003qx}. 

Besides the dark photon dark matter and dark photon superradiance cloud
scenarios covered in this work, similar vortex dynamics may potentially occur in a
gauge boson cosmological collider~\cite{wang2020gauge}, or dark matter
scattering during galaxy mergers~\cite{Lasenby:2020rlf}. In some of these
cases, the dynamics described in this paper, and the subsequent evolution of
the string networks, can lead to other striking signals. We leave these studies
to future work.

\begin{acknowledgments}
We thank Asimina Arvanitaki, Itay Bloch, Neal Dalal, Anson Hook, and Sergey Sibiryakov for helpful discussions. JH would like to also thank Lei Gioia Yang and Liujun Zou for many clarifying discussions about superconductors. Research at Perimeter Institute is supported in part by the Government of Canada through the Department of Innovation, Science and Economic Development Canada and by the Province of Ontario through the Ministry of Colleges and Universities.
WE acknowledge support from an NSERC Discovery grant.  This research was
enabled in part by support provided by SciNet (www.scinethpc.ca) and Compute
Canada (www.computecanada.ca). Calculations were performed on the Symmetry
cluster at Perimeter Institute, the Niagara cluster at the University of
Toronto, and the Narval cluster at Ecole de technologie sup\'erieure in
Montreal.

\end{acknowledgments}

\appendix

\section{Abelian Higgs Simulations} 
\label{sec:num}

As in~\cite{East:2022ppo}, we numerically solve the Abelian-Higgs equations of motion in the Lorenz gauge ($\partial_\mu A'^{\mu}=0$) using $\Phi=\Phi_R+i\Phi_I$, $\partial_t \Phi$, $A'_\mu$, and $E'_i \equiv F'_{it}$ as our
evolution variables. With these variables, the evolution equations take the following form:

\begin{eqnarray}
    \partial_t^2\Phi &=& \partial_i \partial^i \Phi-2i g_D A'^\mu \partial_\mu\Phi -g_D^2A'^{\mu}A'_\mu \Phi-\lambda (|\Phi|^2-v^2)\Phi, \\
\partial_t A'_i &=& -E'_i+\partial_i A'_t,\\
\partial_t A'_t &=& \partial_i A'^i -Z,\\
    \partial_t E'^i &=&  B'^i \tau_a^{-1}(t) +\epsilon^{ijk}\partial_jB_k +g_D^2|\Phi|^2 A'^i  
-g_D\left( \Phi_R \partial^i \Phi_I-\Phi_I \partial^i \Phi_R \right) + \partial^i Z,\\
\partial_t Z &=& 
-\sigma Z + \partial_i E^i -g_D^2|\Phi|^2 A_t 
  +g \Phi_R\partial_t\Phi_I -g\Phi_I\partial_t\Phi_R .
\label{eqn:vec}
\end{eqnarray}

Here, following~\cite{Zilhao:2015tya,East:2017mrj,Helfer:2018qgv}, we introduce an auxiliary variable $Z$ designed to
damp violations of the constraint that the divergence of the electric field
equals the charge density due to numerical truncation error on timescales of $1/\sigma$. 
(That is, in the limit of infinite numerical resolution, $Z$ will be identically zero.) 
We also include the 
source term in the evolution of $E'^i$ that would arise from the coupling 
between the dark photon and axion given by eq.~\ref{eqn:axion_coup}, 
in the case that the axion is spatially homogeneous with $\tau_a(t)=f/\dot{a}$.
For our purposes, we will not directly evolve the axion field, but instead use
the source term as a simplified model of some process that drives the increase in the 
dark photon field for some period of time.

The equations are evolved numerically using similar methods to~\cite{East:2022ppo}. Spatial
derivatives are approximated using fourth order finite differences and the
time integration is performed using fourth order Runge-Kutta. We use a periodic
domain with length $L$ in each spatial direction. In many cases we assume a
translational symmetry in the $z$ direction which allows us to reduce our
computational domain to two dimensions and obtain high resolutions without
prohibitive computational expense.

For the axion source term, we use the time dependence
\begin{equation}
    \tau_a(t)=\bar{\tau}_a I\left(\frac{t-t_0+\Delta t/2}{\Delta t}\right)
\end{equation}
where $I$ smoothly transitions from unity to zero:
\begin{equation}
	I(x)=
\left\{
	\begin{array}{lll}
		1  & \mbox{if } x<0\\
        1-x^3(6x^2-15x+10) & \mbox{if } 0 \leq x \leq 1 \\
		0  & \mbox{if } x>1 \ . \\ 
	\end{array} 
\right. 
\end{equation}

We perform a set of simulations where we fix $\bar{\tau}_a = -0.474/m_{A'}$ and
$L=11.914/m_{A'}$. With these choices, as described in
section~\ref{sec:sim_results}, we expect an exponentially growing vector field, with perturbations of
wavenumber $k_p=4.0\pi/L\approx 1.055 m_{A'}$ growing the fastest, with e-folding
time $\approx 3.0 /m_{A'}$. We start with a small random perturbation in $A'$ 
(and set $\Phi=v$, $\partial_t \Phi=0$ everywhere) and let
the unstable mode grow for a number of e-folds choosing (in most cases) $t_0$ such that the 
instability shuts off shortly after the formation of string vortices, and 
letting $\Delta t\approx 6.0/m_{A'}$. 

\begin{figure}
\begin{center}
\includegraphics[width=0.5\columnwidth,draft=false]{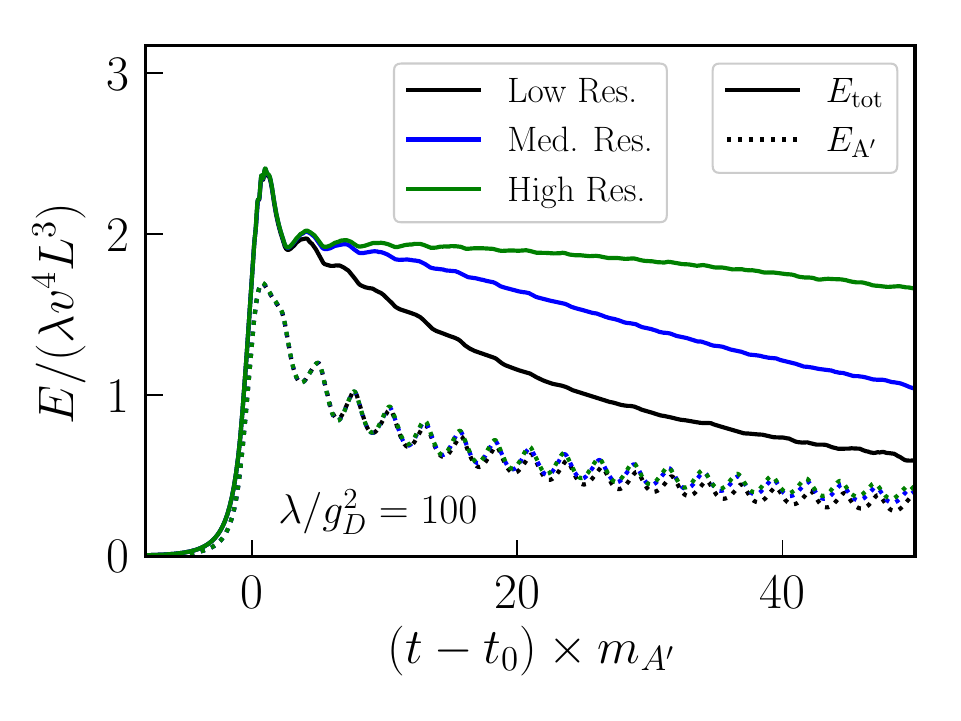}
\end{center}
\caption{
    The total energy (solid lines) and vector field energy (dashed lines) as a function of time for $\lambda/g_D^2=100$, and three different numerical resolutions.
\label{fig:e_conv}
}
\end{figure}

We are interested in the limit of $\lambda \gg
g_D^2$, but this introduces a small scale---the string radius in comparison to 
the wavelength of the exponentially growing mode---which requires increasing resolution.
In figure~\ref{fig:e_conv}, we show a resolution study for a case with $\lambda/g_D^2=100$. 
The highest resolution has $dx\approx 0.06 \lambda^{-1/2}v^{-1}$, while the medium and lower
resolutions have grid spacings that are $2$ and $4\times$ as coarse, respectively. 
As can be seen in the figure, the exponential growth phase is well resolved for all cases, but after
string vortex formation there is a noticeable numerical dissipation in the energy associated with the scalar
sector in the lower resolutions. Unless otherwise stated, we use resolution equivalent to the highest
resolution for all the results presented here. 

We perform simulations with the following parameters. We assume a translation
symmetry in one direction and consider cases with $\lambda/g_D^2=100$, 200, and
400, choosing $t_0$ such that the average energy density reaches $\langle \rho
\rangle \approx 2\lambda v^4$. For $\lambda/g_D^2=100$, we also vary the time of
the axion instability shutoff $t_0$ by $\approx \pm 1.2/m_{A'}$.  Finally, we
consider a fully 3D case assuming no symmetries with $\lambda/g_D^2=25$ and
$L=2\pi/k_p$, as well as the equivalent case with a translational symmetry for
comparison. For this case $dx\approx 0.12 \lambda^{-1/2}v^{-1}$. 

\section{Klein Gordon equation in a background electric and magnetic field}\label{sec:eomKG}

In this appendix, we review the solution of Klein Gordon equation in a background electric and magnetic field. We will work in the Landau gauge and comment on the differences between the electric and magnetic field. Consider the following action
\begin{equation}
    \mathcal{S} = \int {\rm d}^4 x  \left(\left|D_{\mu} \Phi\right|^2 -\frac{1}{4}F^{\mu\nu}F_{\mu\nu}-m^2 \left|\Phi\right|^2\right),
\end{equation}
where $m^2$ can be both positive and negative. We will consider a constant
magnetic field pointing in the z-direction, and an electric field that is either
parallel or perpendicular to the magnetic field. To start with, however, we
consider only a background magnetic field. The correction to the scalar
effective potential from a constant magnetic field is computed
in~\cite{Lam:1971my}. The equation of motion is
\begin{equation}
    \left[ (\vect{p} - e \vect{A})^2 + m^2 \right]\Phi = (p_0- e A_0) \Phi,
\end{equation}
with $A_x = - y B_z$ and $A_0 = A_y=A_z = 0$. The equation of motion can be written as
\begin{equation}
    \left[-\frac{d^2}{d y^2}+(e B_z)^2 (y + p_x/e B_z)^2+ p_z^2 -\omega^2 + m^2  \right] Y[y] = 0,
\label{eq:bfieldbackground}
\end{equation}
where $\Phi = Y[y]\exp [ i p_x x + i p_z z - i \omega t]$. Changing variable from $y$ to $\tilde{y}$ through $\tilde{y} = |e B_z|^{1/2}(y + p_x/eB_z)$, we can rewrite equation \ref{eq:bfieldbackground} as 
\begin{equation}
    \left[-\frac{d^2}{d \tilde{y}^2}+\tilde{y}^2 + \frac{ p_z^2 -\omega^2 + m^2}{e B_z}  \right] Y[y] = 0,
\label{eq:b2}
\end{equation}
and find the solutions with Hermite polynomials $H_n(\tilde{y})$ as $Y (\tilde{y}) = \exp^{-\tilde{y}^2/2} H_n(\tilde{y})$, and the dispersion relation is
\begin{equation}
    \omega^2 = p_x^2 + p_z^2 + m^2 + |e B_z| (2n+1), \quad n = 0,\,1,\,2,\ \ldots,
\end{equation}
which reduces to the standard Landau level result in the non-relativistic limit. A scalar receives a positive correction to its mass squared $m_B^2 = |e B_z| (2n+1)$ in a background magnetic field. Such a correction can help restore a broken $U(1)$ symmetry (the bare $m^2$ is negative) when $|e B_z| >|m^2|$. 

Similarly, we can also add an electric field that is either parallel or
perpendicular to the magnetic field. If the electric and magnetic field are
parallel, we can choose the gauge where $A_x = - y B_z$,  $A_0 = -z E_z$, and
$A_y = A_z = 0$, which has a solution $\Phi = Y[y]Z[z]\exp [ i p_x x - i \omega t]$
satisfying
\begin{align}
    \left[-\frac{d^2}{d y^2}+(e B_z)^2 (y + p_x/e B_z)^2 - m_B^2 \right] Y[y] = 0, \nonumber\\
    \left[-\frac{d^2}{d z^2} - (\omega + e E_z z)^2 + m^2 + m_B^2 \right] Z[z] = 0.
\label{eq:ebackground}
\end{align}
Note that the effect of the electric field and magnetic field is totally separable in this case, and as a result, the solution in a purely electric field background corresponding to solving the equations with $m_B =0$. Performing a similar change of variables $\tilde{z} = (e E_z)^{1/2} (z+\omega/e E_z)$ allows us to rewrite equation \ref{eq:ebackground} as 
\begin{equation}
    \left[-\frac{d^2}{d \tilde{z}^2} -\tilde{z}^2 + \frac{ m_B^2 + m^2}{e E_z}  \right] Y[y] = 0,
\label{eq:e2}
\end{equation}
and find the solution as Kummer functions
\begin{equation}
    Z(\tilde{z}) = F \left[\frac{1}{4} \left(1 + i\frac{m^2+ m_B^2}{4 e E_z}\right),\,\frac{1}{2} ,\, - i \tilde{z}^2\right] \quad \text{or} \quad  F \left[\frac{1}{4} \left(1 + i\frac{m^2+ m_B^2}{4 e E_z}\right)+\frac{1}{2},\,\frac{3}{2} ,\, - i \tilde{z}^2\right].
\end{equation}
Despite the complicated appearance of the solution, the main difference between
the equations of motion in \ref{eq:e2} and \ref{eq:b2} is the sign of the
second term, which determines if the third entry of the Kummer function is real
or imaginary. The solution in the electric field is not an energy eigenstate,
which is expected since charged particles are generically accelerated by the
electric field. Mathematically, both the first and second kind Kummer function
is a viable solution to the equation of motion.  

If the electric field is perpendicular to the magnetic field, we can choose the gauge where $A_x = - y B_z$,  $A_0 = -y E_y$ and $A_y = A_z = 0$. In this case, the electric and magnetic field has effects that are not separable, and the solution is $\Phi = Y[y]\exp [ i p_x x + i p_z z - i \omega t]$, with equation of motion
\begin{equation}
    \left[-\frac{d^2}{d y^2} - e^2 (E_y^2- B_z^2) y^2 - 2 e y (\omega E_y - p_x B_z )  + p_x^2 +p_z^2 - \omega^2 +m^2  \right] Y[y] = 0 .
\end{equation}
For $E_y^2- B_z^2\neq 0$, we can define the Lorentz invariant field strength
$F^2 \equiv E_y^2- B_z^2$, perform a change of variables $\tilde{y} = (e
F)^{1/2}(y - (B_z p_x - \omega E_y)/e F^2)$ and solve the differential
equations. This change of variable reduces to the above-mentioned change of
variables in the limit where $E$ or $B$ is zero. As expected, the coefficient
of the $y^2$ term is Lorentz invariant and gauge invariant, which makes sense,
since if the electric and magnetic field are perpendicular, one can always perform a
Lorentz transformation to go to a frame where only one of the two fields exist.
It should be noted that the above discussion only concerns time-independent and
spatially uniform electric and magnetic fields. The conditions for symmetry
restoration in a background field that is time dependent (radio lasers) and
spatially non-uniform electric field (in particular if $\nabla \times E \neq
0$) is much more complicated, and we refer readers
to~\cite{tinkham2004introduction} for more discussion of this.

To conclude, we showed in this appendix the gauge invariant equations of motion
and eigenfunctions of scalar QED. As we demonstrated with the case including an electric
field, increasing the vector potential $A^{\mu}$, which is not gauge invariant,
does not necessarily contribute to a breaking or restoration of a $U(1)$
symmetry. In the background of a magnetic field, the zero mode of the scalar
field gets a positive mass squared equal to $e B$, and the $U(1)$ symmetry
can be restored globally when $e B > \lambda v^2$, in agreement with the value we
have in equation \ref{eq:condition1}.

\section{Possible resolutions for dark photon production mechanisms}\label{sec:resolution}

Possible resolutions to the problems with the dark photon production
mechanisms implied by vortex formation can be separated into two categories. The first is to add new
dynamics that makes the dark photon mass different during and after
inflation, or more generally, during production and today. Secondly, the
dynamics studied in this paper does not apply to the case where the mass of the
dark photon is introduced simply as a relevant operator that does not have a UV
origin, in which case the mass of the dark photon can be considered as a parameter that
breaks a shift symmetry instead of spontaneous breaking a $U(1)$ gauge
symmetry. In this case, the Goldstone boson that is eaten by the photon to
become the longitudinal polarization of the massive photon is not necessarily
compact, and the dynamics we study in this paper does not happen. In the
following, we outline two model building approaches that can resolve the issues
discussed in this paper following the above-mentioned ideas. 

\subsection{Clockwork Mechanism}\label{sec:clockwork}

\begin{figure}
    \centering
    \includegraphics[width=0.6\textwidth]{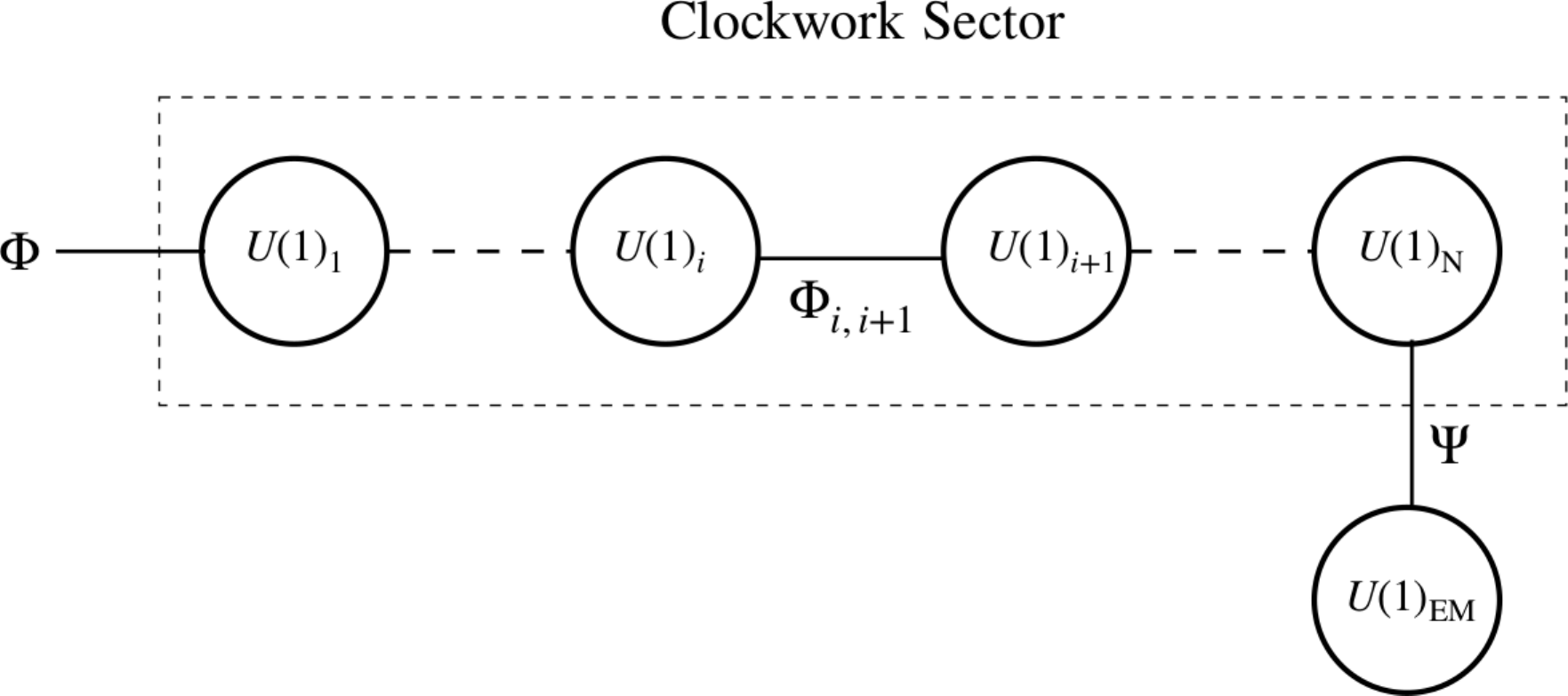}
    \caption{The clockwork model that can generate a large kinetic mixing but a tiny dark photon mass. The various VEVs of scalars $\Phi_{i,\,i+1}$ break the $U(1)^N$ gauge groups in the dashed box down to a single dark $U(1)_D$. The dark photon gets a mass from the VEV of the scalar $\Phi$, while the fermion $\Psi$ generates the kinetic mixing between $U(1)_D$ and $U(1)_{\rm EM}$.}
    \label{fig:GaugeGroup}
\end{figure}

A possible resolution of both the issue of inflationary production, as well as
that of a late time production mechanism, is to introduce a clockwork
mechanism~\cite{Gherghetta:2019coi} that creates a separation between the
charge of the Higgs field and a fermion field $\Psi$ that mediates the mixing
between the dark photon and the standard model photon, so as to have an observable dark photon. Such 
models have been invoked mainly to explain a small kinetic mixing parameter
assuming an $\mathcal{O}(1)$ dark gauge coupling~\cite{Gherghetta:2019coi}. In
our case, we apply the mechanism in the opposite direction to explain the
lightness of the dark photon given a large VEV. Consider the following
Lagrangian
\begin{equation}
    \mathcal{L} = \left|D_{\mu}\Phi_{i,\, i+1}\right|^2 - \frac{\lambda}{4} \left(\left|\Phi_{i,\, i+1}\right|^2 - v_{i,\, i+1}^2\right)^2 -\frac{1}{4} F_{i}^{\mu\nu} F_{i\mu\nu}
\end{equation}
where the covariant derivative $D^{\mu} \Phi_{i,\, i+1} = (\partial^{\mu} - i
q_i g_i A_i^{\mu} + i q_{i+1} g_{i+1} A_{i+1}^{\mu} )\Phi_{ij}$. The gauge
couplings are chosen to be equal ($g_i=g_D$) and the scalar charges are chosen to be
$(q_i , q_{i+1} ) = ( Q_R , 1)$ under the two consecutive gauge groups
$U(1)_i$ and $U(1)_{i+1}$ (see figure~\ref{fig:GaugeGroup}). After symmetry
breaking, $\langle\Phi_{i,\, i+1}\rangle =  v_{i,\, i+1}$, the original
$U(1)^{N}$ gauge symmetry is broken down to a single $U(1)_D$ by the $N-1$
Higgs fields $ \Phi_{i,\, i+1}$, and the Higgs field $\Phi$ and the fermion field
$\Psi$ have an effective charge $(q_{D \Phi},q_{D \Psi}) = (1/Q_R^{N-1} , 1)$
under the remaining gauge group, which we identify as $U(1)_D$. The kinetic
mixing between $U(1)_{\rm EM}$ and $U(1)_{D}$ is generated through a $\Psi$
loop, and current experimental sensitivity probes a region~\cite{Baryakhtar:2017ngi,Cervantes:2022yzp,Chiles:2021gxk,AxionLimits}: 
\begin{equation}
\epsilon = \frac{q_{D\Psi} q_{\Psi} e g_D }{16\pi^2} \gtrsim 10^{-16} \ ,
\end{equation}
while $g_D q_{\Phi} \lesssim 10^{-22}$ from eq.~$\ref{eq:requirement}$. Therefore, for the parameter space of interest, we have 
\begin{equation}
q_{D \Psi}/q_{D \Phi} \gtrsim 10^{8}.
\end{equation}
Such a clockwork mechanism works because in the limit where the gauge group is
$R$ instead of a $U(1)$, the Goldstone direction is no longer compact/periodic,
and hence vortex production is impossible.

However, unlike the normal clockwork models, where the VEVs of the scalar field
that breaks the $U(1)^{N}$ symmetry down to the diagonal can be arbitrary, as
long as they are larger than the VEV that breaks the final $U(1)$ symmetry to
generate a dark photon mass, in the case of late time dark photon production mechanisms,
these VEVs have to be much larger. The clockwork scalars are coupled with up to
$\mathcal{O}(1)$ $U(1)_D$ charges and vortices of these scalars $\Phi_i$ are
more likely to be nucleated in the background field. This is a hierarchy
problem of $v_{i,\, i+1}/v \sim q_{D \Psi}/q_{D \Phi} \sim 10^{8}$.

\subsection{Non-minimal coupling}\label{sec:NonMCoupling}
Another possible resolution to the issue of vortex formation during the inflationary production
of dark photons is to
introduce a non-minimal coupling of the scalar field to the Ricci Scalar $R$
of the form $\xi_{\Phi} |\Phi|^2 R$. Such a coupling can change the potential
of the scalar field. In particular, it can increase the VEV beyond the Hubble
scale when $\xi_{\Phi}<0$ and $|\xi_{\Phi}| \gg 1$, which can suppress 
vortex production. However, such a suppression goes as
\begin{equation}
 \exp [ -4\pi^2 v^2/ H^2] \sim \exp \left [-4\pi^2 \frac{6|\xi_{\Phi}|}{\lambda} \right ].
\end{equation}
for $\lambda \gg 1$, up to power law corrections. This exponential
suppression, however, needs to be compared to the large number of Hubble
patches at the end of inflation, of order $\exp{180}$. If a significant fraction 
of the Hubble patches contain a vortex, these vortices can consume the energy
stored in the dark photon dark matter field produced in the Hubble patches
without a seed vortex and expand, converting a significant portion of the total
energy density into vortices. This suggests that a
large non-minimal coupling might be needed to avoid the system eventually reaching the scaling solution. 

A large non-minimal coupling can help avoid vortex production, however, the
coupling also changes the relation between the dark photon mass and the
inflationary Hubble scale implied by the dark photon dark matter. In
particular, after inflation, both the field $\Pi$ and the Higgs VEV $v$ are
functions of time. The dimensionless angular variable $\Pi/v$, instead of $\Pi$,
will be constant during adiabatic evolution (see the text around
equation~\ref{eq:abelianhiggs} for the definition). In order to not have
strings form due to the Kibble mechanism, it is also required that $\Pi/v$
never exceed $2\pi$.  Therefore, though a large non-minimal coupling can stop
string production during inflation, it still does not help solve the issue of producing 
sufficient dark photon dark matter.

\section{Notation}\label{sec:notation}

\begin{table}[h]
    \centering
    \begin{tabular}{c|c|c|c}
    \hline \hline
        Parameter (DP) & Description (DP) & Parameter (SC) & Description (SC) \\
    \hline
       $ m_{A'}= g_D v$ & dark photon mass & $\delta = m_A^{-1}$ & London penetration depth \\
     $ m_{\rho} = \lambda^{1/2}v $ & dark Higgs mass &$ \xi = m_{\rho}^{-1}$ & coherence length \\
  $ \kappa = \delta/\xi$ &  &  $ \kappa^2 = \lambda/g_D^2$ & \\
    $ \Phi_{0}'= 2\pi/g_D $ & flux quanta &  $ \Phi_{0}= \pi/e$ & flux quanta \\
      $ B_{c1}' = g_D v^2 $ & dark residual field & $ H_{c1} = \frac{\Phi_0}{4\pi \delta^2} \log \kappa $ & first critical field \\
      $ B_{c2}' = \lambda v^2/g_D $ & symmetry restoration & $ H_{c2} = \frac{\Phi_0}{2\pi \xi^2}  $ & second critical field \\
     $ B_{\rm sh}' = \lambda^{1/2} v^2 $ & superheating field & $ H_{\rm sh} = \frac{C_{\rm sh}\Phi_0}{\delta  \xi}  $ & superheating field \\
    \hline \hline
    \end{tabular}
    \caption{A list of useful variables and terminology, their definitions in the dark photon (DP) and superconductor (SC) literature, as well as how the various variables and terminology are related/analogous to each other. A more detailed discussion can be found in~\cite{tinkham2004introduction} and section~\ref{sec:Abrikosov}}.
    \label{tab:variabledefinitions}
\end{table}

\bibliographystyle{JHEP}
\bibliography{reference.bib}

\end{document}